\documentclass{aa}
\usepackage[utf8]{inputenc}
\usepackage{graphicx}
\usepackage{txfonts}
\usepackage[tight,footnotesize]{subfigure}
\usepackage[center]{caption}
\usepackage{multirow}
\usepackage{xcolor}
\usepackage{comment}
\usepackage[flushleft]{threeparttable}
\usepackage{hyperref}
\usepackage{newunicodechar}
\newunicodechar{−}{\ensuremath{-}}
\hypersetup{
    colorlinks=true,
    linkcolor=blue,
    filecolor=magenta,      
    urlcolor=blue,
    citecolor=blue,
}

\begin{document} 
\title{The X-ray and optical emission of the intermediate polar Swift J0614.0$+$1709\thanks{Based on observations obtained with \textit{XMM-Newton}, an ESA science mission with instruments and contributions directly funded by ESA Member States and NASA.}}
\author{ Nikita Rawat\inst{1}\corrauth{rawatn@saao.ac.za},
          Domitilla De Martino\inst{2},
          David A. H. Buckley\inst{1,3,4},
          Koji Mukai\inst{5,6},
          Martina Veresvarska\inst{7},
          Zackery A. Irving\inst{8},
          Simone Scaringi\inst{9},
          Nicola Masetti\inst{10,11}
}

\institute{South African Astronomical Observatory, PO Box 9, Observatory, Cape Town, 7935, South Africa
\and 
Istituto
Nazionale di Astrofisica, Osservatorio Astronomico di Capodimonte, Salita Moiariello 16, I-80131, Napoli, Italy
\email{domitilla.demartino@inaf.it}
\and
Department of Physics, University of the Free State, PO Box 339, Bloemfontein 9300, South Africa
\and
National Astronomical Observatories, Chinese Academy of Sciences,
20A Datun Road, Chaoyang District, Beijing, 1000012, People’s Republic of China
\and
CRESST II and X-ray Astrophysics Laboratory, NASA/GSFC, Greenbelt, MD 20771, USA
\and
Department of Physics, University of Maryland, Baltimore County, 1000 Hilltop Circle, Baltimore, MD 21250, USA
\and
Institute of Space Sciences (ICE, CSIC), Campus UAB, Carrer de Can Magrans s/n, 08193, Barcelona, Spain
\and
School of Physics and Astronomy, University of Southampton, University Road, Southampton SO17 1BJ, UK
\and
Centre for Extragalactic Astronomy, Department of Physics, Durham University, South Road, Durham DH1 3LE, UK
\and
Istituto Nazionale di Astrofisica, Osservatorio di Astrofisica e Scienza dello Spazio di Bologna, Via Gobetti 101, I-40129, Bologna, Italy
\and
Instituto de Astrof\'isica, Facultad de Ciencias Exactas, Universidad Andr\'es Bello, Fern\'andez Concha 700, Las Condes, Santiago, Chile
 }

\renewcommand{\baselinestretch}{1.2}
\newcommand{\orb}{${\Omega}$\,} 
\newcommand{\twoorb}{${2\Omega}$\,} 
\newcommand{\threeorb}{${3\Omega}$\,} 
\newcommand{\po}{P$_{\Omega}$\,} 
\newcommand{\ptwoo}{P$_{2\Omega}$\,} 
\newcommand{\pthreeo}{P$_{3\Omega}$\,} 
\newcommand{\ps}{P$_{\omega}$\,} 
\newcommand{\ptwos}{P$_{2\omega}$\,} 
\newcommand{\pb}{P$_{\omega-\Omega} $\,} 
\newcommand{\ptwob}{P$_{2(\omega-\Omega)} $\,} 
\newcommand{\pbplus}{P$_{\omega+\Omega}$} 
\newcommand{\ptwobplus}{P$_{2(\omega+\Omega)}$} 
\newcommand{\pone}{P$_{\omega-2\Omega}$} 
\newcommand{\ptwo}{P$_{\omega+2\Omega}$} 
\newcommand{\pthree}{P$_{2\omega-\Omega}$} 
\newcommand{\pfour}{P$_{2\omega+\Omega}$} 
\newcommand{\s}{${\omega}$\,} 
\newcommand{\twos}{${2\omega}$\,} 
\newcommand{\be}{$\omega$\,-\,$\Omega$\,} 
\newcommand{\twobe}{$2(\omega$\,-\,$\Omega)$\,} 
\newcommand{\threebe}{3($\omega$\,-\,$\Omega$\,)} 
\newcommand{\beplus}{$\omega$\,+\,$\Omega$\,} 
\newcommand{\sone}{$2\omega$\,-\,$\Omega$\,} 
\newcommand{\stwo}{$2\omega$\,+\,$\Omega$\,} 
\newcommand{\sthree}{$\omega$\,-\,$2\Omega$\,} 
\newcommand{\cts}{Counts s$^{-1}$\,} 

\newcommand{\rn}[1]{%
  \textup{\uppercase\expandafter{\romannumeral#1}}%
}
 
\abstract
{We present the first detailed X-ray and optical study of the intermediate polar Swift J0614.0$+$1709 based on a pointed \textit{XMM-Newton} observation complemented by long-term optical photometry from \textit{TESS}. A strong coherent modulation at the white dwarf spin period was detected in X-rays, with a period of $\sim$1411 s, confirming the magnetic nature of the system. The X-ray spin pulse exhibits a pronounced energy dependence, with the modulation amplitude decreasing towards higher energies, consistent with phase-dependent absorption in the accretion flow. Phase-resolved spectroscopy further shows that the spin variability is primarily driven by changes in the covering fraction of the partial-covering absorber. The broadband X-ray spectrum is well described by a partially absorbed thermal plasma model with a temperature of $\sim$12 keV and a soft blackbody component at $\sim$63 eV. The presence of the soft component places Swift J0614.0$+$1709 among the growing class of `soft' intermediate polars. The inferred bolometric luminosity of $\sim$5$\times$10$^{34}$ erg s$^{-1}$ places the source among the more luminous intermediate polars. The \textit{TESS} observations reveal substantial sector-to-sector variability in the relative strengths of the orbital and spin modulations. While the spin signal remains persistently dominant, the beat modulation varies significantly and disappears entirely in one Sector, suggesting temporal changes in the contribution of the reprocessing region(s). The orbital and, to a lesser extent, spin modulation amplitudes show some tendency to be stronger during brighter optical states, suggesting that changes in the mass transfer rate may influence the optical emission and modulation amplitudes. Overall, Swift J0614.0$+$1709 is a comparatively X-ray luminous intermediate polar exhibiting complex optical variability characterised by changes in pulse morphology and recurrent brightening events, making it a valuable target for future simultaneous optical and X-ray studies. }


\titlerunning{Swift J0614.0$+$1709}
\authorrunning{N. Rawat et al.}
\maketitle
\nolinenumbers

\section{Introduction}
Cataclysmic variables (CVs) comprise a population of compact interacting binaries in which mass transfer onto a white dwarf (WD) occurs via Roche-lobe overflow from a typically late-type main-sequence donor \citep[see][for a comprehensive review]{1995cvs..book.....W}. These systems are commonly divided into magnetic and non-magnetic classes, depending on the extent to which the accretion flow is governed by the WD's magnetic field.  In magnetic CVs, at some point the accretion proceeds under the influence of the WD's magnetic field, which guides the inflowing material along field lines onto the magnetic poles. The flow attains supersonic velocities before being abruptly decelerated in a standing shock above the WD surface. The post-shock region reaches temperatures of $\sim$10--50 keV and cools via thermal bremsstrahlung, producing hard X-ray emission, as well as via polarised cyclotron radiation at optical and near-infrared wavelengths \citep{1973PThPh..49.1184A}, whose relative importance depends on the magnetic field strength and mass accretion rate \citep{1996A&A...306..232W}. In addition, part of the high-energy emission is reprocessed in the WD atmosphere, while diamagnetic `blobs' in the accretion flow can also penetrate the WD photosphere, both of which can give rise to a softer blackbody-like component observable in the soft X-ray and/or EUV/UV bands. The hard X-rays can also be reflected from the WD surface, producing a Compton reflection component as well as a fluorescent Fe feature at 6.4 keV \citep{mukai2015}.

\par Magnetic CVs are further subdivided according to the field strength of the WD. Polars represent the more strongly magnetised subclass (B $\gtrsim$10 MG) in which the magnetic field is sufficiently intense to inhibit the formation of an accretion disc \citep[see][for a review]{1990SSRv...54..195C}. In these systems, the WD rotation is synchronised, or very nearly so, with the binary orbit such that the spin period of the WD (\ps) equals the orbital period of the binary system (\po).  Intermediate polars (IPs), in contrast, host WDs with generally weaker magnetic fields (B $\approx$1--10 MG), allowing the formation of a truncated accretion disc, although a few systems appear to be discless \citep[e.g. V2400 Oph;][]{1997MNRAS.287..117B}. These systems are asynchronous, with \ps $<$ \po \citep[see][for a review of IPs]{1983ASSL..101..155W, 1994PASP..106..209P, 1995ASPC...85..185H}. 

\par The X-ray power spectra of IPs often exhibit multiple periodic components, including spin, beat, orbital, and other sideband frequencies, reflecting the coupling between the rotating magnetosphere and the binary frame. Because the X-ray emission arises directly in the post-shock region, the relative prominence of these signals depends on how material is transferred to the WD. Consequently, the relative strengths of the spin, beat, and sideband signals observed in X-rays are widely used to trace the underlying accretion geometry. Accretion onto the WD in IPs is commonly described in terms of three modes, disc-fed, stream-fed, and disc-overflow, determined by system parameters such as the magnetic field strength of the WD, mass accretion rate, and orbital separation. In disc-fed systems, an accretion disc forms but is truncated at the magnetospheric radius, from where the accretion flow is channelled along magnetic field lines, forming accretion curtains that direct material onto the WD poles \citep{1988MNRAS.231..549R}. These systems predominantly exhibit variability at the spin frequency (\s) of the WD \citep{1995A&A...298..165K, 1996MNRAS.280..937N}. In contrast, when the magnetic field is sufficiently strong to inhibit disc formation, accretion proceeds directly via a stream that couples to the magnetosphere and is channelled onto the magnetic pole caps, a scenario referred to as stream-fed (or disc-less) accretion \citep{1986MNRAS.218..695H}. In such cases, variability is commonly observed at the beat frequency (\be), reflecting the changing geometry between the accretion flow and the rotating WD \citep{1991MNRAS.251..693H, 1992MNRAS.255...83W, 1993MNRAS.265..316N}. \citet{1992MNRAS.255...83W} showed that additional modulation at the spin frequency can arise even in stream-fed accretion if the two magnetic pole caps are non-identical, differing in size, shape, or luminosity. Disc-overflow accretion \citep{1989ApJ...340.1064L, 1996ApJ...470.1024A} corresponds to an intermediate regime in which a truncated disc coexists with a stream component that bypasses the disc and couples directly to the WD magnetosphere \citep{1989MNRAS.238.1107H, 1991ApJ...378..674K}. This scenario naturally gives rise to variability at both the spin and beat frequencies \citep[see][]{1991MNRAS.251..693H, 1993MNRAS.265L..35H}.  Although the same periodicities are often detected at optical wavelengths, their interpretation is more complicated because a substantial fraction of the optical emission may arise from reprocessing and geometric effects. In several well-studied systems, such as TX~Col and FO~Aqr, long-term and time-resolved observations show that the dominant frequency components can change with time, indicating changes in the dominant accretion mode \citep[see][for details]{1993MNRAS.265L..35H, 1996rftu.proc..123B, 1997MNRAS.289..362N, 1999ASPC..157...47W,2007MNRAS.380..133M,2020ApJ...896..116L, 2021ApJ...912...78R}.

\par Swift J0614.0$+$1709 (hereafter J0614) was listed as a CV candidate in the \textit{Swift}-BAT 70-month hard X-ray survey \citep{baumgartner2013}. It was first studied optically by \citet{2015AJ....150..170H}, who identified properties suggestive of a magnetic CV. Their timing analysis revealed three prominent peaks in the power spectrum at 5.00(5) h, 1412.3(9) s, and 1529.8(9) s, which were interpreted as the orbital, spin, and beat periods, respectively. The optical spectrum showed Balmer emission lines and He \rn{2} 4686 \AA  ~comparable in strength to H$\beta$, supporting a magnetic classification. \citet{hou2023} further investigated the optical properties of the source using LAMOST spectroscopy and \textit{TESS} data, obtaining results consistent with those of \citet{2015AJ....150..170H}. The \textit{TESS} light curves were analysed in detail by \citet{bruch2025}, who reported variations in the accretion flow structure inferred from changes in the relative power of the spin and beat frequencies. In addition, \citet{suleimanov2025S} estimated a WD mass of 0.37$\pm$0.25 M$_\odot$ based on the \textit{Swift}-BAT spectrum. Despite these extensive optical investigations, many aspects of its X-ray behaviour remain poorly characterised, leaving a key aspect of the system unexplored. In this work, we address this gap by presenting detailed analysis of pointed \textit{XMM-Newton} observations of J0614 complemented by long-term \textit{TESS} data, enabling a comprehensive study of its X-ray and optical emission properties, including both variability and spectral characteristics.

\par The paper is organised as follows. In Section \ref{sec2}, we summarise the observations and data reduction. Sections \ref{sec3} and \ref{sec4} present the analysis and results of the X-ray and optical data. Finally, Sections \ref{sec5} and \ref{sec6} provide the discussion and conclusions, respectively.

\begin{figure*}
\centering
\subfigure[]{
\includegraphics[width=16cm, height=6cm]{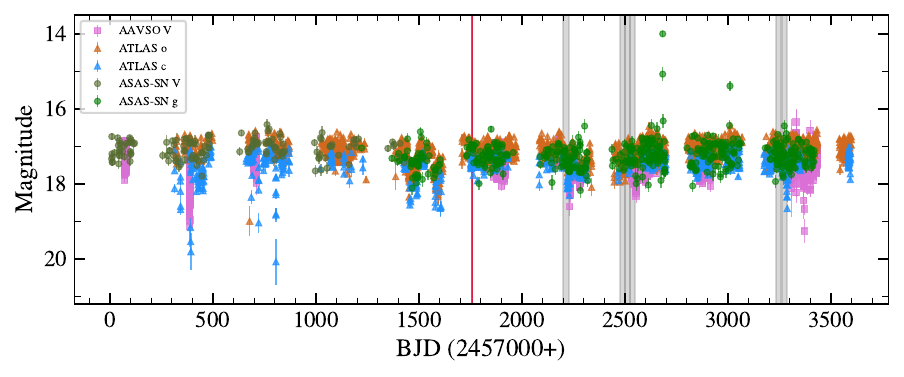} \label{fig:long_term_lc}}
\subfigure[]{
\includegraphics[width=15.5cm, height=8cm]{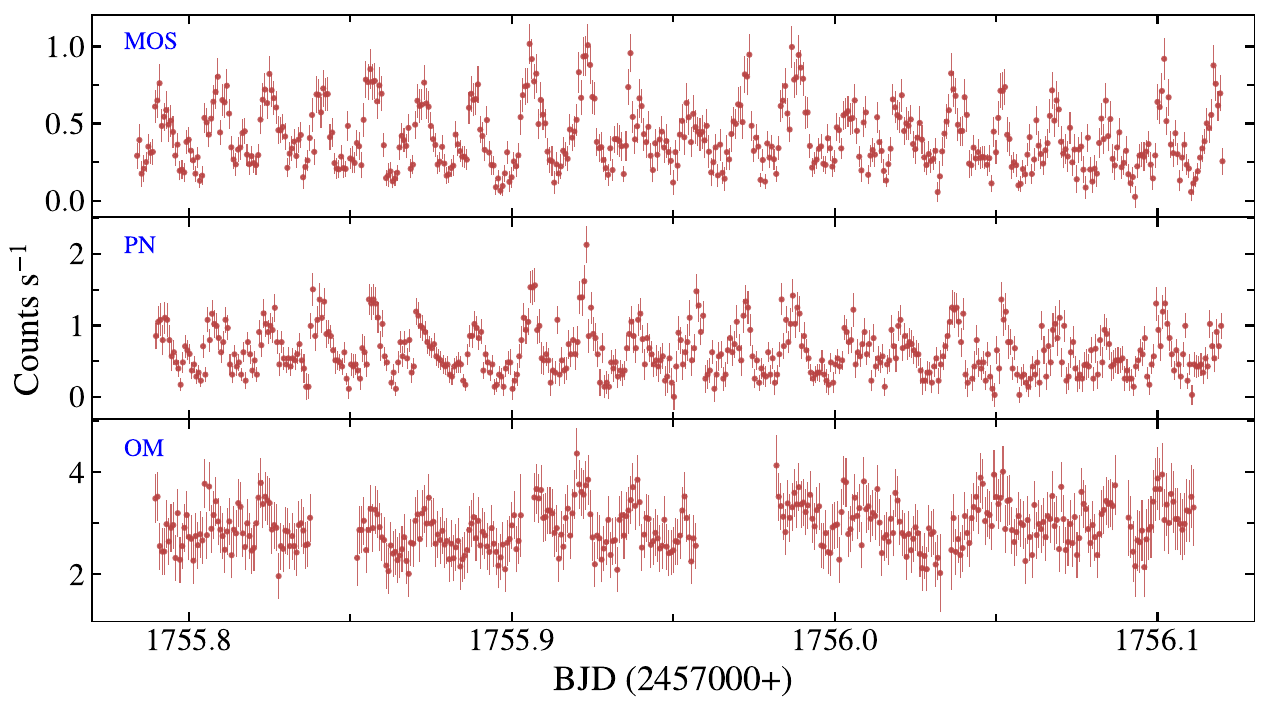} \label{fig:lc_xmm}}
\subfigure[]{
\includegraphics[width=15.5cm, height=7cm]{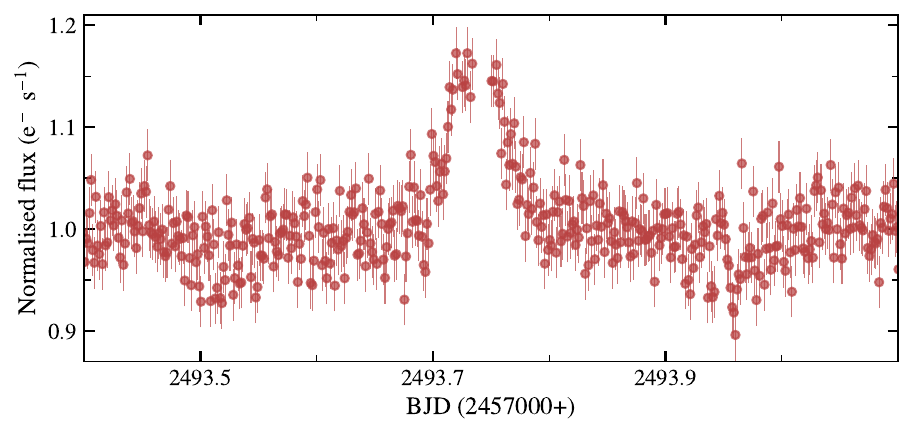} \label{fig:lc_tess}}

\caption{(a) Combined long-term AAVSO, ASAS-SN, and ATLAS light curve of J0614. The time span of the \textit{XMM-Newton} and \textit{TESS} observations is indicated by the red- and grey-shaded regions, respectively. (b) EPIC light curves in the 0.2–12.0 keV energy range and OM light curve in B filter, binned in 60 s intervals. (c) Representative segment of the \textit{TESS} light curve from Sector 43 showing the brightening event described in Section \ref{sec3.1}. } 
\label{fig:lc}
\end{figure*}

\section{Observations and data reduction} \label{sec2}
\subsection{\textit{XMM-Newton} observation}
The \textit{XMM-Newton} observation of J0614 (observation ID: 0841180101) was performed on 29 September 2019, starting at 06:44:40.323 UTC, with the European Photon Imaging Camera \citep[EPIC;][]{2001A&A...365L..18S, 2001A&A...365L..27T} operated in small-window mode with the thin filter \citep{2001A&A...365L...1J}. The net exposure times amount to 29.2 ks (EPIC-MOS) and 28.7 ks (EPIC-PN). Data reduction was carried out with the standard \textit{XMM-Newton} \textsc{science analysis system (sas)} software package (version 22.1.0) using the calibration files available on 20 May 2025.\footnote{\url{https://www.cosmos.esa.int/web/xmm-newton/current-calibration-files}} The EPIC event files were processed with \textit{epproc} and \textit{emproc} within \textsc{sas}, following the standard threads.\footnote{\url{https://www.cosmos.esa.int/web/xmm-newton/sas-threads}} Photon arrival times were then barycentre-corrected using the task \textit{barycen}. We found no evidence of high background flaring activity in the data. The presence of pile-up was investigated using \textit{epatplot} and found to be insignificant. Events were extracted from a $30^{\prime\prime}$ circular region around the source, with the background taken from a nearby region of identical extent free of sources. The spectral response files were generated using \textit{rmfgen} and \textit{arfgen}, and the spectra were grouped to a minimum of 20 counts per bin using \textit{specgroup} in order to apply $\chi^2$ statistics.  The light curves and spectra were obtained over the 0.2--12.0 keV and 0.3--10.0 keV energy range, respectively. The MOS1 and MOS2 light curves were co-added using \textit{lcurve}. The RGS data were also inspected but not analysed further because the source was not significantly detected, resulting in insufficient signal-to-noise for meaningful high-resolution spectroscopy. Further temporal and spectral analysis was carried out with \textsc{heasoft} (version 6.35.1).

\begin{figure*}[h!]
\centering
\subfigure[]{\includegraphics[width=8cm, height=8.5cm]{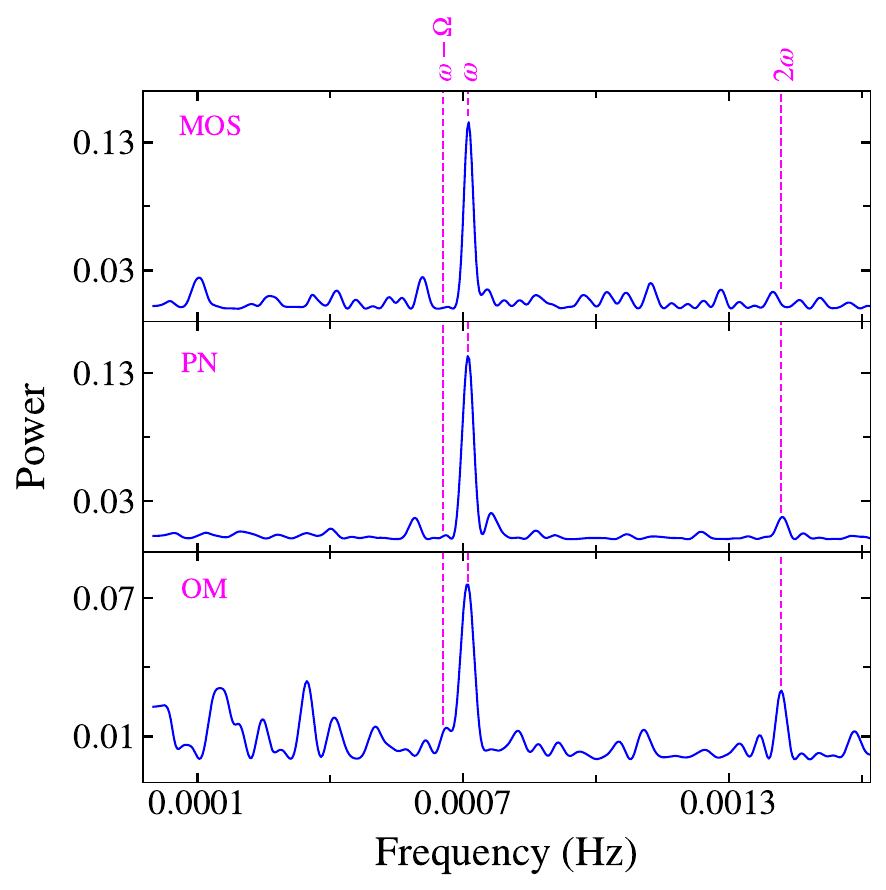} \label{fig:xmm_ps}}
\subfigure[]{\includegraphics[width=8cm, height=9cm]{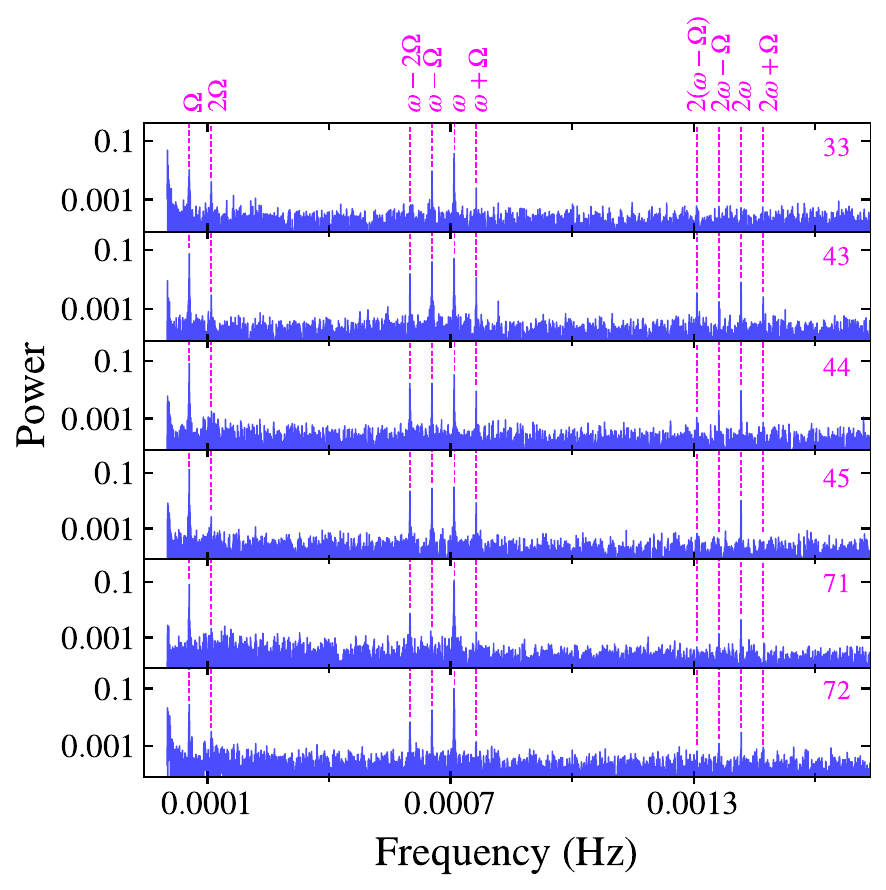} 
\label{fig:tess_ps}}
\caption{Power spectra of J0614 derived from (a) \textit{XMM-Newton} and (b) \textit{TESS} data, with the dominant frequencies marked for visual clarity.}
\label{fig:ps}
\end{figure*}

\par  J0614 was also observed with the Optical Monitor \citep[OM;][]{2001A&A...365L..36M} on board \textit{XMM-Newton}, with a total exposure of 28 ks obtained in fast imaging mode using the B filter (3900–4900 $\AA$). The fast-mode data were reprocessed with the \textit{omfchain} pipeline. The six individual exposures were barycentric corrected and subsequently combined to produce a single summed light curve. 

\subsection{\textit{TESS}, AAVSO, ASAS-SN, and ATLAS observations}
The \textit{TESS} observations of J0614 were obtained between 18 December 2020 and 7 December 2023 in Sectors 33, 43, 44, 45, 71, and 72, and they comprise all available 2-min cadence data for the source. The \textit{TESS} bandpass spans 600–1000 nm, with an effective wavelength of 800 nm \citep[see][for details]{2015JATIS...1a4003R}. The data were retrieved from the Mikulski Archive for Space Telescopes (MAST)\footnote{\url{https://mast.stsci.edu/portal/Mashup/Clients/Mast/Portal.html}} under the unique identifier `434329100'.  We primarily used the simple aperture photometry (SAP) flux provided by the \textit{TESS} pipeline, retaining only cadences with a quality flag value of zero. The light curves were detrended by fitting a linear trend independently to each half of a sector and normalising the fluxes by the resulting model. 

\par We also utilised publicly available data of J0614 from the American Association of Variable Star Observers (AAVSO) in the V band, the All-Sky Automated Survey for Supernovae \citep[ASAS-SN;][]{2014ApJ...788...48S, hart2023asassnskypatrolv20} in the V band, and the Asteroid Terrestrial-impact Last Alert System \citep[ATLAS;][]{2018PASP..130f4505T, 2018AJ....156..241H, 2020PASP..132h5002S, 2021TNSAN...7....1S} in the o and c bands. These data were used to investigate the long-term behaviour of the source.

\section{Timing analysis} \label{sec3}
\subsection{Light curves and power spectra} \label{sec3.1}
The long-term optical light curve of J0614 is shown in Fig.~\ref{fig:long_term_lc}, with the source observed at different brightness levels over time. The background-subtracted, combined MOS, PN, and OM light curves exhibit a clear and coherent periodic variability on short timescales (see Fig.~\ref{fig:lc_xmm}). The mean net count rates are $0.33$ counts s$^{-1}$ for the MOS and $0.53$ counts s$^{-1}$ for the PN. The OM light curve yields an average net count rate of $2.90 \pm 0.02$ counts s$^{-1}$, corresponding to an instrumental magnitude of $18.110 \pm 0.007$ mag.\footnote{We adopted a zero point of 19.2661 for the B filter, as provided in the \textit{XMM-Newton} Users Handbook (\url{https://xmm-tools.cosmos.esa.int/external/xmm_user_support/documentation/uhb/XMM_UHB.pdf}).} The Lomb–Scargle power spectra of all light curves were computed using the \textsc{astropy} package,\footnote{\url{https://docs.astropy.org/en/stable/index.html}} which is well suited to unevenly sampled data and naturally accommodates any data gaps in the observations. The resulting power spectra for the MOS, PN, and OM data are presented in Fig.~\ref{fig:xmm_ps}. The dominant peaks in each dataset correspond to periods of $1407.7 \pm 1.8$ s, $1407.2 \pm 2.4$ s, and $1410.2 \pm 7.1$ s for the MOS, PN, and OM, respectively. All three values are mutually consistent within their uncertainties, differing by less than 0.5$\sigma$. This period is identified as the spin period of the WD, previously reported from optical observations (e.g. \citealt{2015AJ....150..170H, bruch2025}), and is here detected for the first time in the X-ray band. Additional support for this interpretation is provided by the independent \textit{TESS} timing analysis presented below, which recovers both the spin and beat frequencies. The uncertainties on the derived periods were estimated using a bootstrap approach, following the methodology described in \citet{irving2024} and \citet{paice2024} and are quoted at the 1$\sigma$ confidence intervals. We note that the beat frequency is not clearly resolved in the power spectrum and the orbital period cannot be derived from the short \textit{XMM-Newton} observation.

\par The \textit{TESS} light curves of J0614 have been analysed in detail by \citet{bruch2025}. We complemented their study with a sector-by-sector Lomb--Scargle analysis, providing an independent assessment of the periodic signals and their stability across different sectors. The resulting power spectra are shown in Fig.~\ref{fig:tess_ps}, where the dominant frequencies are indicated. We identified the frequencies \orb, \twoorb, \s, \twos, \be, \twobe, \beplus, \sone, \stwo, and \sthree\ across the dataset, and their corresponding periods are listed in Table~\ref{tab:periods_tess}. Although the detected frequencies broadly agree with those reported by \citet{bruch2025}, our sector-by-sector analysis shows that the presence and relative strengths of the periodic signals vary between sectors, a behaviour that is not apparent when multiple sectors are combined. In particular, the \be\ frequency is not detected in Sector~71, although it becomes apparent when Sectors~71 and 72 are analysed together \citep{bruch2025}. The $\sim$24 d duration of the Sector~71 observations spans approximately 1359 beat cycles, indicating that the non-detection is unlikely to arise from insufficient temporal coverage. By combining all sectors, we derive an orbital period of $5.09215 \pm 0.00002$ h, a spin period of $1411.718 \pm 0.003$ s, and a beat period of $1529.477 \pm 0.005$ s. As for the \textit{XMM-Newton} data, the uncertainties on the derived periods were estimated using a bootstrap approach. These orbital and beat period values were subsequently adopted for folding all light curves discussed in the following section.
\par In addition to the periodic variability, \citet{bruch2025} reported a brightening event in \textit{TESS} Sector~43. In Fig.~\ref{fig:lc_tess}, we present a representative segment of the Sector~43 light curve showing this event. The long-term ASAS-SN light curve further reveals at least two additional occasions on which the source brightened by approximately 2--3 mag (see the ASAS-SN data points in Fig.~\ref{fig:long_term_lc}), displaying similar transient behaviour. Based on the limited number of detected events, we place a tentative upper limit of $\sim$200--300 days on the recurrence timescale. These observational signatures suggest that the events may be qualitatively similar either to micronova-like bursts \citep{scaringi2022} or to bursts associated with magnetically gated accretion reported in other systems \citep[][and references therein]{ilkiewicz2024}; however, a detailed investigation of this possibility is beyond the scope of the present work. We also examined the available optical survey data obtained within 2 min of the \textit{XMM-Newton} observation and found no evidence of similar brightening events. Consequently, the \textit{XMM-Newton} data are unlikely to have been affected by such events. 

\begin{figure}[h!]
\centering
\includegraphics[width=8cm, height=12cm]{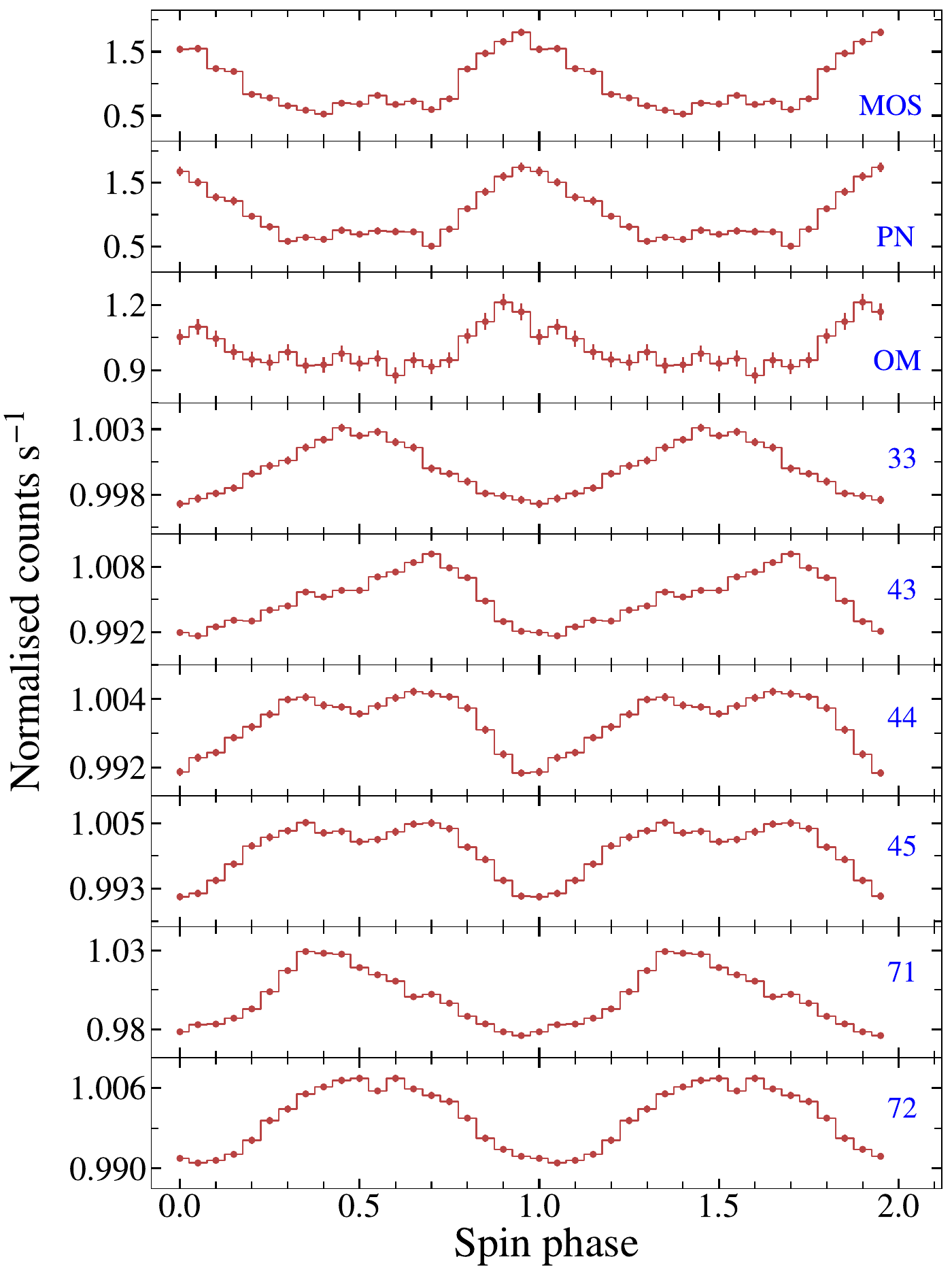} 
\caption{Light curves folded at the spin period for the MOS, PN, OM, and \textit{TESS} datasets with a phase binning of 20. }
\label{fig:folded_tess_xmm}
\end{figure} 

\begin{table*}
\centering
\caption{Modulation amplitudes for the orbital, spin, and beat periods derived from the folded light curves.}
\label{tab:pulse-fraction-tess}
\renewcommand{\arraystretch}{1.2}
\begin{tabular}{lcccccc}
\hline
\multirow{2}{*}{Dataset} &
\multicolumn{2}{c}{Orbital amplitude (\%)} &
\multicolumn{2}{c}{Spin amplitude(\%)} &
\multicolumn{2}{c}{Beat amplitude(\%)} \\
\cline{2-7}
& Measured & Corrected & Measured & Corrected & Measured & Corrected \\
\hline
PN & --- & --- & $55 \pm 4$ & --- & --- & --- \\
OM & --- & --- & $16 \pm 3$ & --- & --- & --- \\
33 & $0.24 \pm 0.02$ & $6.0 \pm 0.5$ &
     $0.29 \pm 0.02$ & $7.2 \pm 0.5$ &
     $0.13 \pm 0.02$ & $3.2 \pm 0.5$ \\
43 & $0.98 \pm 0.06$ & $10.8 \pm 0.7$ &
     $1.00 \pm 0.06$ & $11.0 \pm 0.7$ &
     $0.77 \pm 0.06$ & $8.5 \pm 0.7$ \\
44 & $1.01 \pm 0.05$ & $12.8 \pm 0.6$ &
     $0.72 \pm 0.05$ & $9.1 \pm 0.6$ &
     $0.53 \pm 0.05$ & $6.7 \pm 0.6$ \\
45 & $1.29 \pm 0.05$ & $13.5 \pm 0.5$ &
     $0.68 \pm 0.05$ & $7.1 \pm 0.5$ &
     $0.62 \pm 0.05$ & $6.5 \pm 0.5$ \\
71 & $2.60 \pm 0.11$ & $14.4 \pm 0.6$ &
     $2.66 \pm 0.11$ & $14.8 \pm 0.6$ &
     $0.54 \pm 0.11^{\dagger}$ & $3.0 \pm 0.6^{\dagger}$ \\
72 & $0.76 \pm 0.05$ & $8.8 \pm 0.6$ &
     $0.84 \pm 0.05$ & $9.7 \pm 0.6$ &
     $0.30 \pm 0.05$ & $3.5 \pm 0.6$ \\
\hline
\end{tabular}

\bigskip
\textbf{Notes.}  The corrected amplitudes represent approximate intrinsic amplitudes after accounting for flux dilution.\\
\textbf{$^\dagger$} The beat amplitudes for sector 71 are given as upper limits.
\end{table*}

\subsection{Ephemeris and folded light curves}
To examine the long-term phase stability of the spin modulation, we derived a linear ephemeris using the minima timings measured from the folded \textit{TESS} light curves of the six individual sectors. A linear fit between the cycle numbers and minima timings provided the following ephemeris for J0614:
\begin{equation} \label{eqn}
T_{0} = 2459201.7407(2) + 0.01633932(1) \times E, 
\end{equation}
where $T_{0}$ is defined as the time of minima and the errors are given in parentheses. The spin period in equation (\ref{eqn}) is equivalent to  1411.7172 $\pm$ 0.0009 s. Propagating the ephemeris uncertainties over the $\sim$450 days interval between the first \textit{TESS} sector and \textit{XMM-Newton} observations yields an accumulated phase uncertainty of only $\sim$0.02 cycles, indicating that the cycle count remains uniquely preserved over this baseline.

\par In Fig.~\ref{fig:folded_tess_xmm}, we present the  \textit{XMM-Newton} light curves together with the \textit{TESS} datasets, folded at the spin period using the ephemeris given in equation (\ref{eqn}). The simultaneous OM light curve is broadly consistent with the X-ray light curves, with the principal maxima and minima occurring at comparable spin phases. There is also a subtle hint of a secondary maximum in the MOS and PN light curves around phase $\sim$0.5, approximately half a cycle away from the primary maximum. In contrast, the \textit{TESS} spin profiles show substantial variations in morphology between sectors. The minima occur near the same spin phase in all \textit{TESS} sectors, whereas the maxima are more complex and exhibit significant sector-to-sector variability, ranging from single-peaked to double-humped structures. In addition, the \textit{TESS} light curves appear to show a phase offset of approximately 0.5 cycles relative to the \textit{XMM-Newton} light curves. We measured the pulse fraction of the spin modulation, defined as $(\rm I_{\rm max} - \rm I_{\rm min})/(\rm I_{\rm max} + \rm I_{\rm min}) \times 100$ per cent, where $\rm I_{\rm max}$ and $\rm I_{\rm min}$ are the maximum and minimum count rates of the folded profiles. The amplitude of the modulation exhibits significant variability across the \textit{TESS} sectors (see Table~\ref{tab:pulse-fraction-tess}), reaching a minimum of 0.29\% in Sector~33 and a maximum of 2.66\% in Sector~71. Moreover, the amplitudes of the \textit{TESS} modulations are relatively smaller than those observed in the OM B-band light curves, possibly due to dilution from nearby sources arising from the large pixel scale of \textit{TESS}. We identified three Gaia sources within $\sim$42$^{\prime\prime}$ of the target in the \textit{TESS} target pixel file, including two relatively bright sources with  $G\approx$15.0 and $\approx$14.3 mag and a fainter source with $G\approx$17.9 mag. We investigated the variability of these nearby sources using available ASAS-SN and ATLAS photometry and found no evidence of significant photometric variability, indicating that any variable contamination from these sources is negligible. Their essentially constant flux can nevertheless dilute the observed modulation amplitudes.

\par To quantify the level of flux dilution, we examined the crowding parameter, CROWDSAP, provided by the \textit{TESS} pipeline, which estimates the fraction of the total flux within the photometric aperture that is contributed by the target. We found that CROWDSAP varies substantially between sectors (0.04--0.18), reflecting changes in the sector-dependent photometric apertures and consequently, in the degree of flux dilution. The observed spin modulation amplitudes are strongly correlated with CROWDSAP, indicating that variations in flux dilution are a major contributor to the apparent sector-to-sector amplitude variations. As an illustrative exercise, we divided the observed spin modulation amplitudes by the corresponding CROWDSAP values to estimate the effect of flux dilution. Although this correction is only approximate, the corrected amplitudes still exhibit sector-to-sector variations, suggesting that intrinsic changes in the spin modulation amplitude may also be present. The corrected amplitudes are listed in Table \ref{tab:pulse-fraction-tess}, where the quoted uncertainties include only the statistical uncertainties of the measured amplitudes.

\par
\begin{figure}[h!]
\centering
\includegraphics[width=8cm, height=8cm]{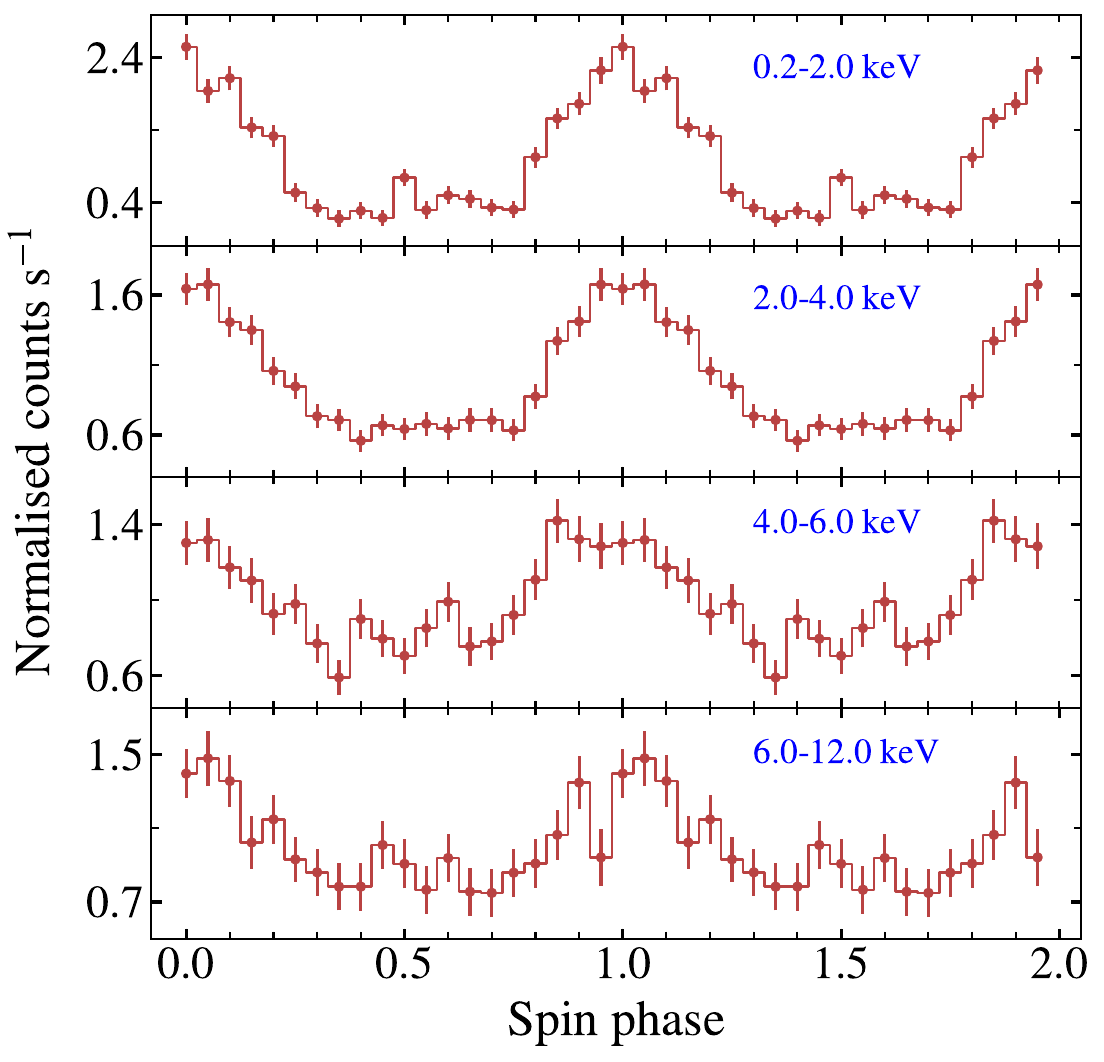} 
\caption{Energy-dependent folded light curves of PN data at spin period with a phase binning of 20.}
\label{fig:folded_ed}
\end{figure}

\begin{figure}[h!]
\centering
\includegraphics[width=8cm, height=7cm]{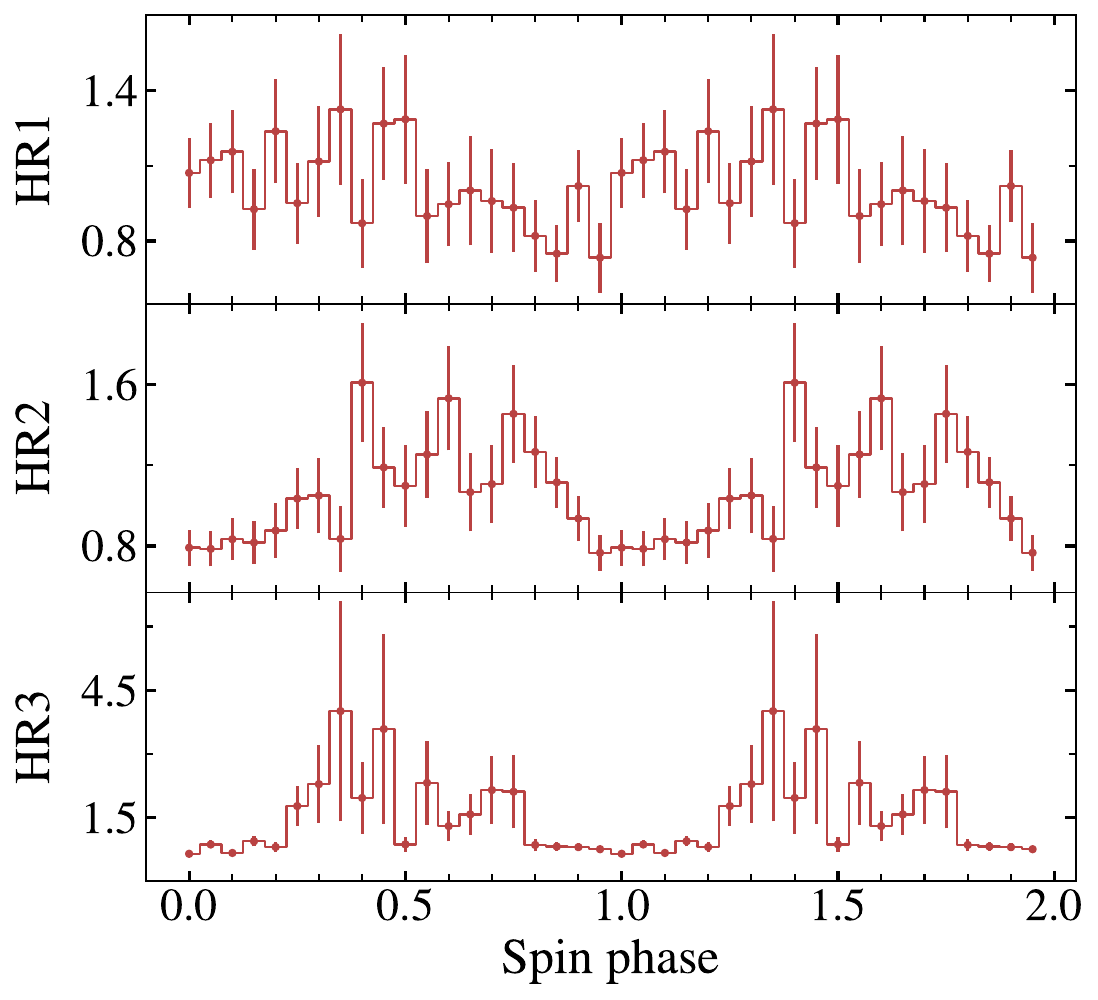} 
\caption{Hardness ratio curves HR1, HR2, and HR3 of PN data with a phase binning of 20. The definitions of the hardness ratios are given in the text.}
\label{fig:hr}
\end{figure}

 Moreover, to investigate the energy dependence of the spin modulation, we folded the energy-resolved PN light curves in the bands 0.2–2.0 keV, 2.0–4.0 keV, 4.0–6.0 keV, and 6.0–12.0 keV using the ephemeris given in equation (\ref{eqn}) (see Fig. \ref{fig:folded_ed}). 
The pulse fraction was calculated using the same definition as above, and the resulting modulation amplitudes are listed in Table~\ref{tab:pulse-fraction}. A clear energy dependence is observed, with the spin modulation becoming more pronounced at lower energies. This behaviour suggests that photoelectric absorption within the accretion flow, commonly observed in IPs, plays a significant role in shaping the observed variability. We further examined the spectral variability through hardness ratio (HR) curves folded at the spin period (Fig.~\ref{fig:hr}). We define three hardness ratios: HR1 = (6–12)/(4–6), HR2 = (4–6)/(2–4), and HR3 = (2–4)/(0.2–2), where the values represent count rate ratios between adjacent energy bands. The HR curves show clear modulation over the spin cycle, particularly in HR2 and HR3, which vary approximately in anti-phase with respect to the intensity profiles. This anti-phase behaviour suggests that the spectral variability is driven by phase-dependent absorption, with enhanced absorption at flux minima preferentially reducing the soft X-ray emission and thereby producing a harder spectrum.

\begin{table}[h]
\centering
\caption{Energy-dependent modulation amplitudes measured from the folded PN light curves at the spin period.}
\label{tab:pulse-fraction}
\renewcommand{\arraystretch}{1.4}
\begin{tabular}{ccc}
\hline
Energy bands (keV) && Amplitude (\%) \\
\hline
0.2-2.0  && 87 $\pm$ 8 \\
2.0-4.0  && 50 $\pm$ 6  \\
4.0-6.0  && 41 $\pm$ 7  \\
6.0-12.0  && 33 $\pm$ 9  \\
\hline
\end{tabular}
\end{table}

\par
\begin{figure*}
\centering
\subfigure[]{\includegraphics[width=8cm, height=8.5cm]{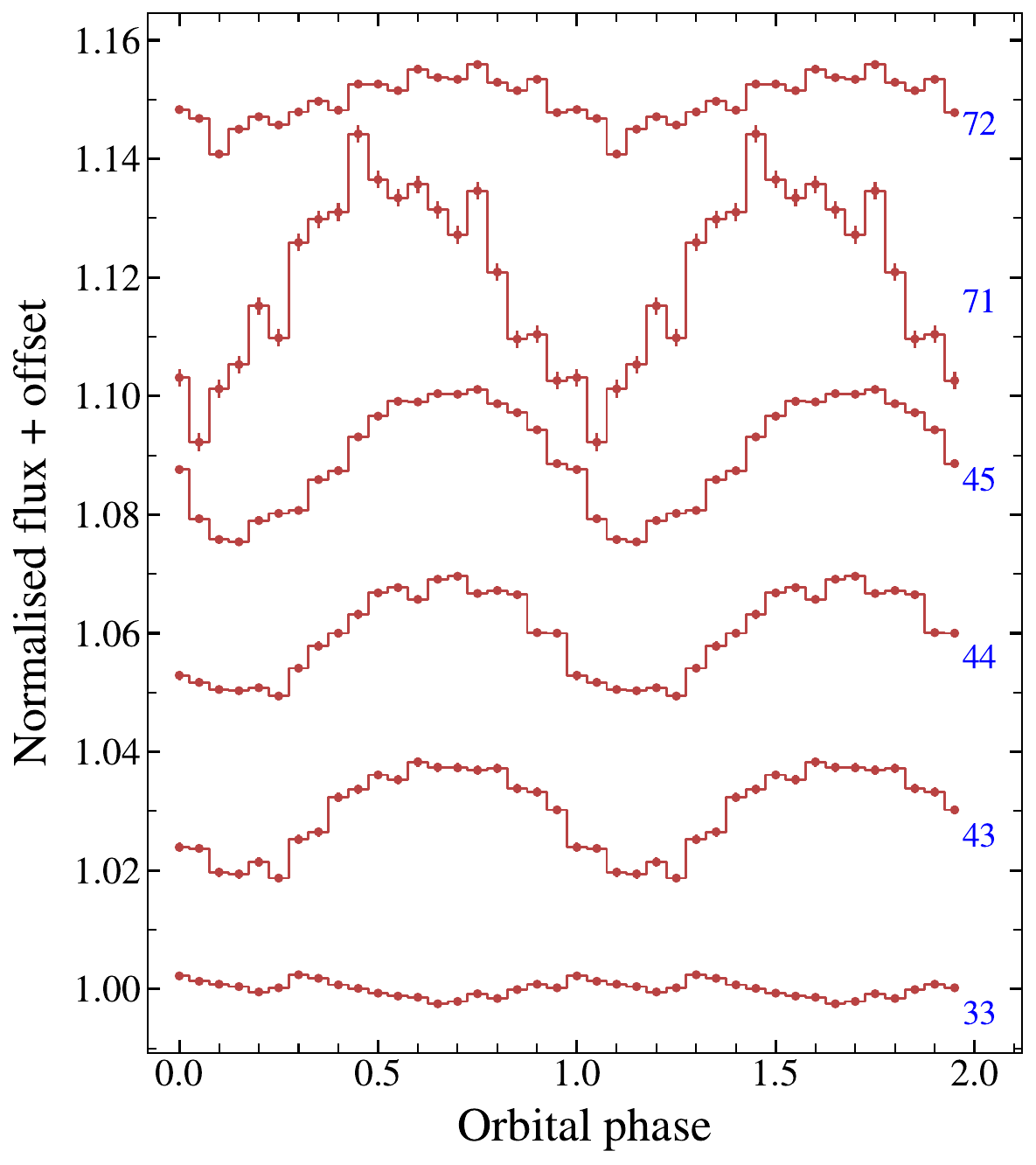} \label{fig:folded_orb}}
\subfigure[]{\includegraphics[width=8cm, height=8.5cm]{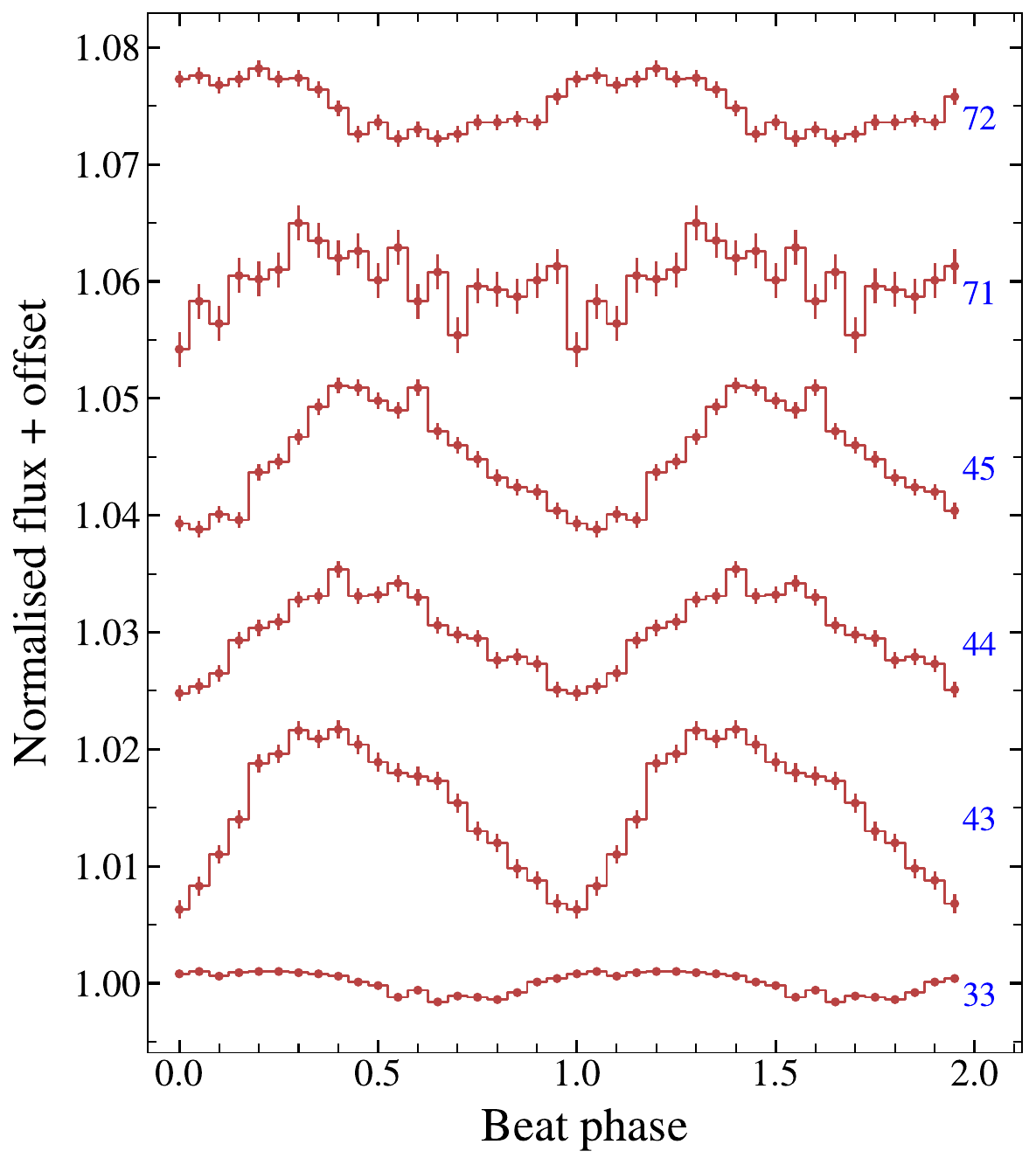} \label{fig:folded_beat}}
\caption{Light curves folded at the orbital and beat periods for the \textit{TESS} datasets with a phase binning of 20. For clarity, successive sectors have been vertically offset in normalised flux by 0.03 and 0.015 for the orbital and beat profiles, respectively.}
\label{fig:folded_tess}
\end{figure*}

 We folded the \textit{TESS} datasets at the orbital and beat periods derived in Section~\ref{sec3.1}, adopting the same reference epoch as used for the spin folding (Fig.~\ref{fig:folded_tess}). The observed orbital modulation amplitudes vary significantly across the \textit{TESS} sectors (Table~\ref{tab:pulse-fraction-tess}), ranging from a minimum of 0.24\% in Sector~33 to a maximum of 2.60\% in Sector~71. After applying the same approximate correction for flux dilution, the orbital modulation amplitudes range from $\sim$6\% to $\sim$14\%. On the other hand, the observed beat modulation is systematically weaker than both the orbital and spin modulations, with amplitudes ranging from a minimum of 0.13\% in Sector~33 to a maximum of 0.77\% in Sector~43. The approximately corrected beat modulation amplitudes range from $\sim$3\% to $\sim$9\% (Table~\ref{tab:pulse-fraction-tess}). The corrected orbital and beat modulation amplitudes therefore still exhibit appreciable sector-to-sector variations across the \textit{TESS} sectors.

\begin{figure*}[h!]
\centering
\subfigure[]{\includegraphics[width=8cm, height=7cm]{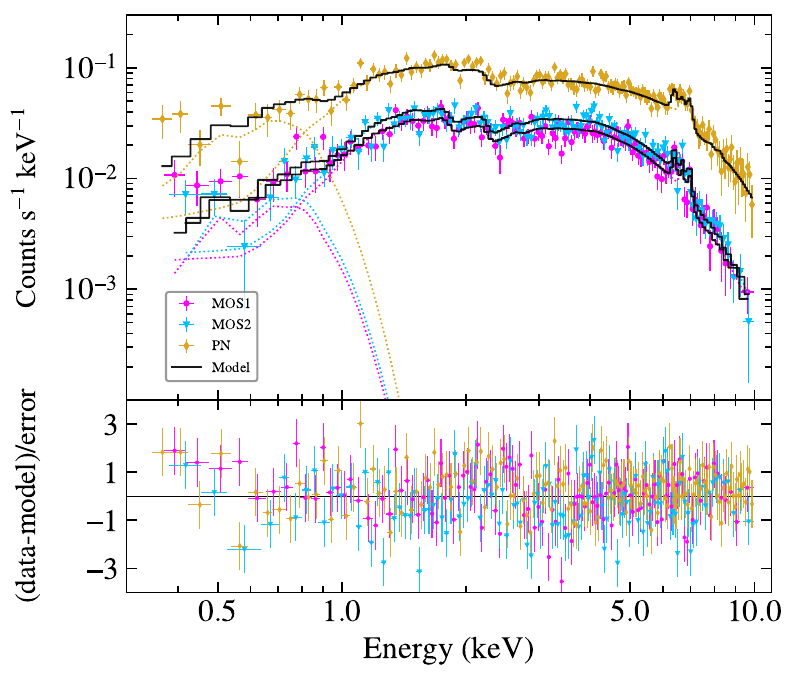} \label{fig:avg_spec}}
\subfigure[]{\includegraphics[width=8cm, height=7cm]{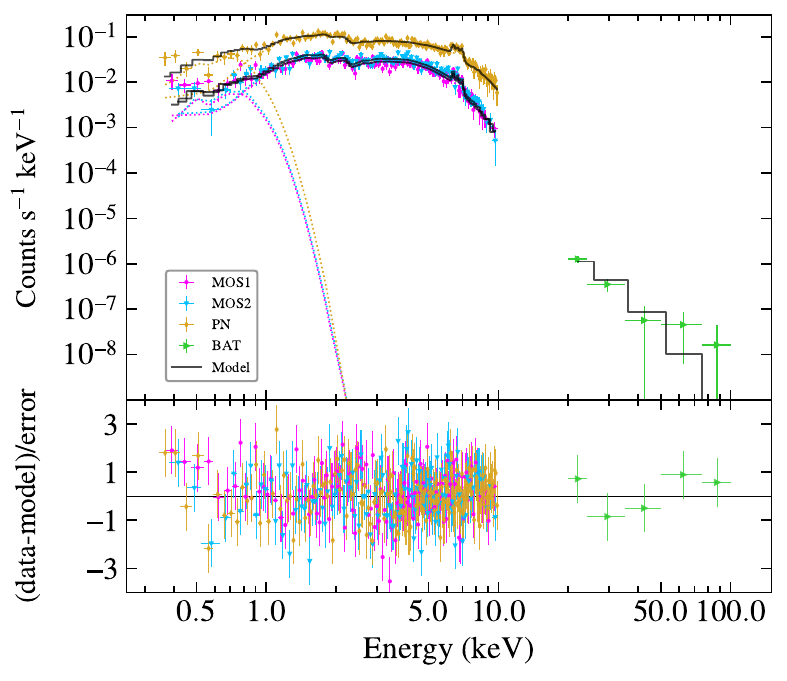} 
\label{fig:avg_spec_bat}}
\caption{(a) MOS1, MOS2, and PN spectra of J0614 fitted simultaneously with the best-fitting model. (b) Same as (a) but including BAT data (see Section \ref{sec4.1} for details). The bottom panel shows the $\chi^{2}$ contribution of data points for the respective model in terms of residual.}
\label{fig:avg_spec_all}
\end{figure*}

\section{Spectral analysis} \label{sec4}
\subsection{Time-averaged spectral modelling} \label{sec4.1}
The background-subtracted X-ray spectra of J0614 (see Fig.~\ref{fig:avg_spec}) were examined over the 0.3–10.0 keV energy range using \textsc{xspec} version 12.15.0 \citep{1996ASPC..101...17A, 2001ASPC..238..415D}. 
In general, the X-ray spectra of IPs are described by multi-temperature collisionally ionised thermal plasma emission originating in the post-shock region, modified by both interstellar absorption and intrinsic partial covering absorbers. To model the observed spectrum of J0614, we employed a combination of standard components available in \textsc{xspec} and the best-fitting parameters were derived using the $\chi^2$ statistic.
\par We started with an initial model that included \texttt{tbabs} to account for interstellar absorption along the line of sight, along with the astrophysical plasma emission code \citep[\texttt{apec};][]{smith2001, foster2012}, which describes emission from collisionally ionised plasma associated with the accretion flow. The \texttt{apec} component is parameterised by the plasma temperature (T$_{\rm apec}$, in keV), metal abundance relative to solar (Z), and normalisation (N$_{\rm apec}$, in units of cm$^{-5}$). The normalisation corresponds to the volume emission measure and is given by $\frac{10^{-14}}{4\pi D^2}\int n_e n_H dV$, where $D$ is the distance to the source, and $n_e$, $n_H$, and $V$ represent the electron density, hydrogen density, and emitting volume, respectively. We adopted the abundance values of \cite{2000ApJ...542..914W} in our fits. 

\par Since a fraction of the hard X-ray emission is expected to be reflected from the WD surface and/or from the pre-shock material \citep[see][]{1999ApJS..120..277E} and the presence of an emission feature near 6.4 keV suggests that such a reflection component may be present, we convolved the thermal plasma emission with the \texttt{reflect} model.  The \texttt{reflect} model is parameterised by the reflection scaling factor, redshift, elemental abundance, iron abundance, and inclination angle. We tied the elemental abundance of the \texttt{reflect} component to that of the \texttt{apec} component, fixed the redshift at zero, the iron abundance at the solar value, and the inclination angle at the default value (cos $i$ =0.45). The reflection scaling factor was fixed at unity, corresponding to the standard isotropic-source approximation. This model provided a poor fit to the data ($\chi^2_\nu > 2$) and was therefore rejected.

\par Residuals at lower energies suggested the presence of additional absorption, as also indicated by the energy dependence of the spin pulsation. We therefore included the convolution model \texttt{partcov} together with \texttt{tbabs} to account for partial covering absorption. The \texttt{partcov} component introduces a covering fraction parameter (cf) describing the fraction of emission affected by the absorber. A Gaussian component with its line energy fixed at 6.4 keV is also included to account for fluorescent Fe K$\alpha$ emission. The Gaussian line width was initially allowed to vary but remained unconstrained by the data. We verified that fixing the width between 0.01 and 0.05 keV yielded consistent fit statistics and best-fitting parameters, whereas larger widths caused the line to become increasingly blended with the underlying continuum. We therefore adopted a fixed line width of 0.02 keV for all subsequent fits.

\par Although the inclusion of partial-covering absorption substantially improved the fit ($\chi^2_{\nu}$= 1.10), the plasma temperature remained pegged at the upper model limit and the elemental abundance was poorly constrained. We therefore explored the inclusion of an additional \texttt{apec} component. Since the fit did not converge when both plasma temperatures were allowed to vary freely, we fixed one temperature at a representative value of 12 keV (or \(0.06\) keV for the alternative fit; see Table \ref{tab:spec_fitting}). Neither model improved the fit statistics relative to the single-temperature model and the other plasma temperature remained pegged at the upper model limit. We also tested a \texttt{mkcflow} component to account for multi-temperature plasma emission, but this fit likewise failed to converge. Consequently, the present \textit{XMM-Newton} data do not uniquely constrain the underlying plasma temperature distribution.

\par The remaining soft X-ray residuals motivated us to explore the inclusion of a blackbody component (\texttt{bb}). The blackbody component is characterised by a temperature (T$_{\rm bb}$, in keV) and normalisation (N$_{\rm bb}$) given by $\rm R_{\rm km}^2/\rm D_{10}^2$, where $\rm R_{\rm km}$ is the emitting radius in km and $\rm D_{10}$ is the distance in units of 10 kpc. The addition of this component significantly improved the fit and yielded well-constrained parameters. We therefore adopted `\texttt{tbabs$\times$(partcov$\times$tbabs)$\times$(bb+reflect$\times$apec+gauss)}' as the final best-fitting model, with a reduced chi-square of $\chi_\nu^2 = 1.05$.

\begin{table*}
\renewcommand{\arraystretch}{1.3}
	\centering
\caption {Spectral parameters obtained from fitting the average spectra of J0614 using best-fitting model= \texttt{`tbabs$\times$(partcov$\times$tbabs)$\times$(bb+reflect$\times$apec+gauss)}'.} \label{tab:spec_fitting}
\begin{tabular}
{p{0.4\columnwidth}p{0.5\columnwidth}p{0.25\columnwidth}p{0.25\columnwidth}}
\hline
Model & Parameters  &  without BAT & with BAT\\
\hline
tbabs & N$_{\rm H}$ ($10^{22} \rm ~cm^{-2}$) & $1.0_{-0.2}^{+0.2}$ & $1.0_{-0.2}^{+0.2}$ \\
partcov &  cf  & $0.73_{-0.04}^{+0.04}$ & $0.73_{-0.04}^{+0.04}$\\
tbabs & N$_{\rm H,cf}$ ($10^{22} \rm ~cm^{-2}$) & $8.4_{-1.4}^{+1.5} $  & $8.5_{-1.3}^{+1.4} $\\

\multirow{2}{*}{bb} & T$_{\rm bb}$ (eV)  &  $63.0_{-6.3}^{+7.4} $  & $62.8_{-6.3}^{+7.4} $\\
& N$_{\rm bb}$ (10$^{-3}$)   &  $1.8_{-1.2}^{+3.8} $ & $1.8_{-1.2}^{+3.6} $ \\

\multirow{3}{*}{apec} & T$_{\rm apec}$ (keV)  &  $12.3_{-1.8}^{+14.0} $ & $11.8_{-1.5}^{+3.1} $ \\
& Z (Z$_{\odot}$) & $0.7_{-0.2}^{+0.5} $ &  $0.7_{-0.2}^{+0.2} $\\
& N$_{\rm apec}$ (10$^{-3}$)   &  $3.9_{-0.4}^{+0.3} $ & $3.9_{-0.3}^{+0.2} $\\

\multirow{2}{*}{gauss} & F$_{\rm 6.4}$ (10$^{-6}$)   &  $7.6_{-2.2}^{+2.2} $ & $7.6_{-2.2}^{+2.2} $ \\
& EW$_{6.4}$ (eV) &  $106_{-27}^{+31} $ & $110_{-19}^{+19} $\\

\multirow{1}{*}{Unabsorbed X-ray flux} & F$_{\rm X, unabs}$  (10$^{-11}$ erg cm$^{-2}$ s$^{-1}$)    &    $4.97_{-0.07}^{+0.07}$ & ---  \\

\multirow{1}{*}{Absorbed X-ray flux} & F$_{\rm X, abs}$  (10$^{-12}$ erg cm$^{-2}$ s$^{-1}$)    &    $4.60_{-0.06}^{+0.06}$ & ---  \\

\multirow{1}{*}{Bolometric flux} & F$_{\rm X, bol}$ (10$^{-10}$ erg cm$^{-2}$ s$^{-1}$)    &    $1.60_{-0.02}^{+0.02}$  &  $1.68_{-0.02}^{+0.02}$\\
\multirow{1}{*}{Bolometric luminosity} & L$_{\rm X, bol}$ (10$^{34}$ erg s$^{-1}$)    &    $5.11_{-0.07}^{+0.07}$  & $5.35_{-0.07}^{+0.07}$ \\
& $\chi_\nu^2$ (dof)                       &  1.05 (400) & 0.99 (402)          \\
\hline
\end{tabular}

\bigskip
\textbf{Notes.} EW$_{6.4}$ and F$_{6.4}$ are the equivalent width and Gaussian line flux, respectively, with F$_{6.4}$ in units of photons $\mathrm{cm^{-2} ~s^{-1}}$; F$_{\rm X, unabs}$ and F$_{\rm X, abs}$ are unabsorbed and absorbed X-ray flux in the 0.3-10 keV energy band, respectively; F$_{\rm X, bol}$ is the unabsorbed bolometric flux derived using a dummy response over the 0.001–100.0 keV energy range; L$_{\rm X, bol}$ is the corresponding bolometric luminosity calculated using the \textit{Gaia} distance of 1630$_{-192}^{+270}$ pc \citep{2021AJ....161..147B}.
\end{table*}

\par To extend the spectral analysis to higher energies, we also included the average spectrum from the BAT 157-month catalogue \citep{lien2025},\footnote{\url{https://swift.gsfc.nasa.gov/results/bs157mon/324}} along with the corresponding response files.\footnote{\url{https://swift.gsfc.nasa.gov/results/bs157mon/}} We note that no significant contamination from nearby sources is reported for J0614 in the BAT 157-month survey catalogue. The BAT spectrum was analysed in its standard distributed survey format with associated statistical uncertainties. We carried out a joint EPIC/BAT spectral fit over the broader 0.3–75.0 keV energy range, using the $\chi^2$ statistic to test whether the extended energy coverage provides additional constraints on the high-temperature end of the post-shock plasma. We therefore tested both a two-temperature model, comprising two \texttt{apec} components, and a cooling-flow model, \texttt{mkcflow}. The two-temperature model improved the fit relative to the single-temperature model and yielded temperatures of $\sim$94 eV and $\sim$11.5 keV for the cooler and hotter components, respectively. 
The normalisation of the cooler component was poorly constrained, with a best-fitting value of 0.49 and large uncertainties, implying an unrealistically large emitting region. We therefore regard this component as physically implausible. Furthermore, the \texttt{mkcflow} model did not converge to a stable solution. Thus, the available data do not provide robust constraints on the temperature distribution of the post-shock plasma. We therefore retained our previously adopted best-fitting model for the final joint fit (see Fig.~\ref{fig:avg_spec_bat}). We further estimated the unabsorbed bolometric flux in the 0.001–100.0 keV band by incorporating the \texttt{cflux} component into the model. The spectral parameters obtained from the simultaneous fitting of the MOS, PN, and BAT spectra are listed in Table \ref{tab:spec_fitting}, with uncertainties quoted at the 90\% confidence level for a single parameter.

\begin{figure*}[h!]
\centering
\includegraphics[width=16cm, height=7cm]{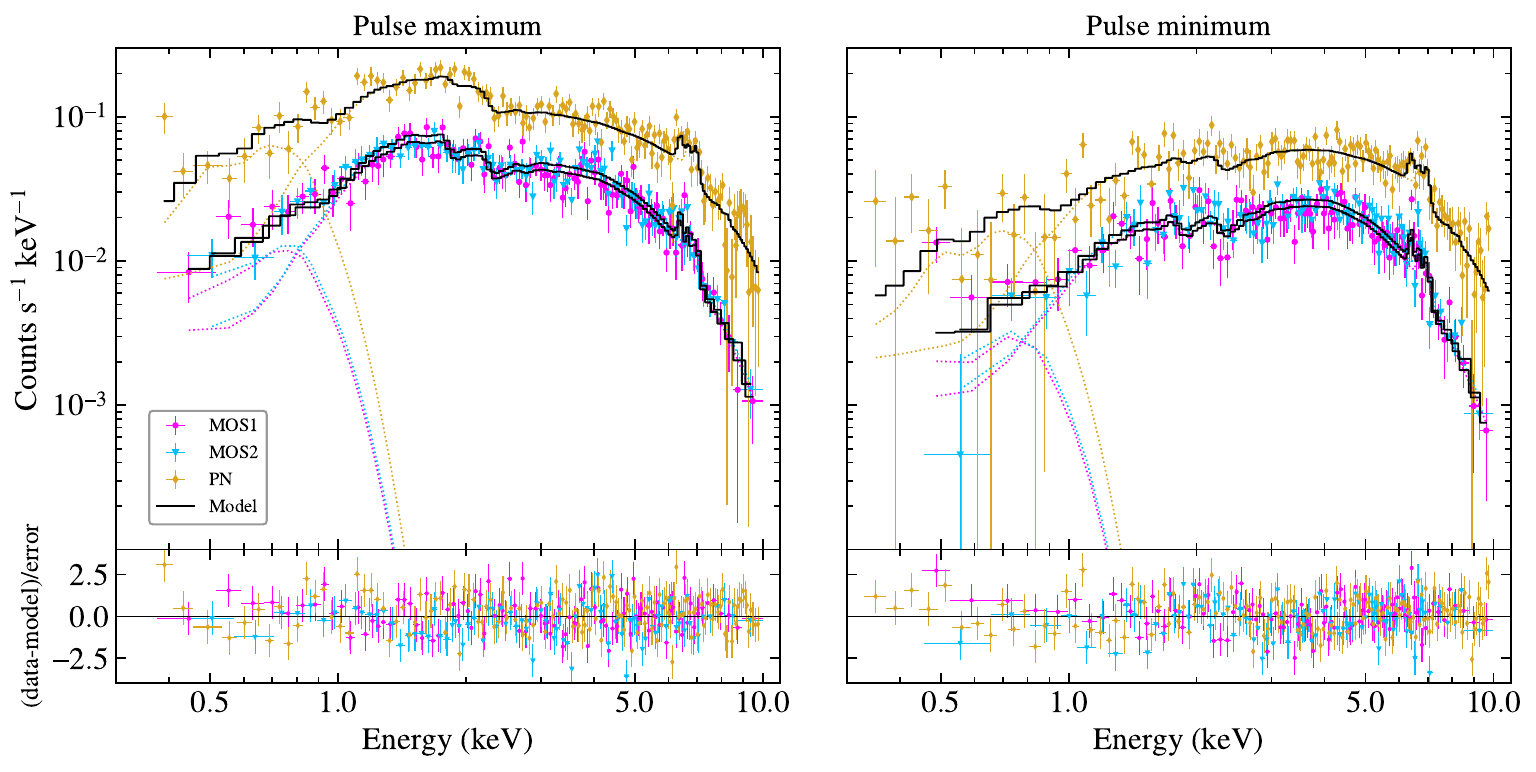} 
\caption{Phase-resolved spectra at the pulse maximum and pulse minimum, fitted with the best-fitting model (see Section \ref{sec4.2} for details). The bottom panel in each case shows the residuals.}
\label{fig:phase_resolved_spectra}
\end{figure*} 

\subsection{Phase-dependent spectral variation} \label{sec4.2}
To investigate possible variations in the spectral parameters across the spin cycle, we extracted MOS and PN spectra corresponding to the pulse maximum and minimum, as identified from the folded light curve (see top two panels of Fig.~\ref{fig:folded_tess_xmm}). The phase intervals were defined as 0.2–0.8 for the minimum and 0.8–1.0 together with 0.0–0.2 for the maximum. The spectra were fitted using the best-fitting model obtained from the fit to the average spectrum, employing the $\chi^2$ statistic. All parameters were fixed to their time-averaged values, except for the covering fraction, N$_{\rm H,cf}$, and the normalisations of the thermal and Gaussian components, which were allowed to vary. We also allowed the \texttt{apec} temperature to vary as a consistency check and found it to remain consistent between the maximum and minimum phases, as well as with the value derived from the average spectral fit, within uncertainties. The phase-resolved spectra, along with the best-fitting model, are shown in Fig. \ref{fig:phase_resolved_spectra}, and the resulting spectral parameters are listed in Table~\ref{tab:phase_resolved}, with uncertainties quoted at the 90\% confidence level for a single parameter.

\par The phase-resolved spectral analysis shows that the covering fraction varies significantly between the pulse maximum and minimum, while the column density remains constant within $\sim$1$\sigma$. The observed X-ray flux also displays a highly significant modulation, consistent with the strong spin variability seen in the light curve. Further, the normalisation of the optically thin plasma component varies considerably ($\sim$5$\sigma$). As for the soft blackbody component, the much larger uncertainties preclude us from making a definitive statement. The iron fluorescence line properties, including both flux and equivalent width, remain consistent within uncertainties, implying that the line-emitting region is largely unaffected by the spin-phase dependent absorption. 

\par To test whether changes in the intrinsic emission, such as self-occultation of the emission region, could be responsible for the spin modulation, we fitted the phase-resolved spectra using two alternative models. The first model corresponds to a pure self-occultation scenario in which only the visible emitting area changes. We therefore employed the model `\texttt{tbabs$\times$(bb+reflect$\times$apec+gauss)}', fixing all parameters to their time-averaged values and allowing only the blackbody and thermal plasma normalisations to vary. The second model assumes a constant partial-covering absorber throughout the spin cycle. Accordingly, we employed the model `\texttt{tbabs$\times$(partcov$\times$tbabs)$\times$(bb+reflect$\times$apec+gauss)}', fixing the covering fraction and N$_{\rm H,cf}$ to their time-averaged values and allowing only the blackbody and thermal plasma normalisations to vary. Both models yielded significantly poorer fits than the model in which the partial-covering absorber was allowed to vary. At the pulse minimum, the $\chi^2_{\nu}$ (dof) values were found to be 7.2 (329) and 2.1 (329) for the first and second models, respectively, whereas at the pulse maximum, the corresponding values were found to be 3.0 (339) and 1.5 (339). These results demonstrate that changes in the intrinsic emission alone cannot account for the observed variability, as also expected from the energy-dependent spin amplitudes. Thus, while self-occultation may contribute to the spin modulation, it is unlikely to be its dominant cause.

\begin{table}
\renewcommand{\arraystretch}{1.3}
	\centering
\caption {Spectral parameters as derived from fitting the EPIC MOS and PN spectra at maximum and minimum spin phases using the best-fitting model. } \label{tab:phase_resolved}
\begin{tabular}{p{0.28\columnwidth}p{0.28\columnwidth}p{0.28\columnwidth}}
\hline
Parameters  &  Pulse maximum & Pulse minimum \\
\hline

cf  & $0.56_{-0.02}^{+0.02}$ & $0.86_{-0.01}^{+0.01}$\\
N$_{\rm H,cf}$ ($10^{22} \rm ~cm^{-2}$) & $7.2_{-1.2}^{+1.4} $  &  $9.0_{-0.9}^{+1.0} $ \\

N$_{\rm bb}$ (10$^{-3}$)   &  $2.4_{-0.4}^{+0.4} $  & $1.9_{-0.7}^{+0.7} $\\

N$_{\rm apec}$ (10$^{-3}$)   &  $4.5_{-0.2}^{+0.2} $  & $3.4_{-0.1}^{+0.1} $\\

F$_{\rm 6.4}$ (10$^{-6}$ )   &  $8.8_{-3.6}^{+3.6} $ &  $7.2_{-2.7}^{+2.7} $\\

EW$_{6.4}$ (eV) &  $106_{-40}^{+38} $ & $114_{-50}^{+50} $\\

F$_{\rm X, unabs}$ (10$^{-11}$erg~cm$^{-2}$~s$^{-1}$) & $6.6_{-0.1}^{+0.1}$ & $5.2_{-0.1}^{+0.1} $\\ 

$\chi_\nu^2$ (dof)   &  1.12 (336)      &   1.11 (326)  \\
\hline
\end{tabular}

\end{table}

\section{Discussion} \label{sec5}
Using X-ray and optical timing analyses together with X-ray spectral modelling, we investigated the variability and accretion properties of the IP J0614. These are discussed in detail in the following sections.

\subsection{Origin of the spin variability in X-rays}\label{sec5.1}
X-ray variability at the spin frequency is a defining characteristic of IPs and can arise from two primary mechanisms: (i) phase-dependent absorption in the accretion flow, including photoelectric absorption and electron scattering \citep{1988MNRAS.231..549R}, and (ii) self-occultation of the X-ray emitting regions by the WD \citep{king1984, 1995A&A...298..165K}. In both cases, the observed emission may originate from one or both magnetic poles, leading to single- or double-peaked spin profiles depending on the system geometry. 

\par The X-ray variability of J0614 is dominated by spin modulation, as evident from the power spectral analysis, which shows the spin frequency and a weaker first harmonic. This is reflected in the X-ray spin profiles that exhibit a weakly double-peaked morphology, with a subtle secondary maximum that could reflect a contribution from the second magnetic pole, while the substantially stronger primary maximum indicates the dominance of the primary accreting pole during the \textit{XMM-Newton} observation. An interesting comparison is provided by the IP GK Per, which also exhibits dwarf-nova-like properties and showed transitions between single- and double-peaked spin profiles during its 2023 outburst. These transitions were interpreted in terms of changes in the visibility of the accretion regions and the accretion geometry \citep{kimura2025}.

For J0614, the phase-resolved X-ray spectroscopy indicates that the modulation is primarily driven by variations in partial-covering absorption, consistent with the standard accretion curtain model for IPs \citep{1989MNRAS.237..853N}. Specifically, the covering fraction varies significantly between pulse maximum and pulse minimum, whereas the column density remains comparatively stable.
As the WD rotates, the observer sees changing aspects of the accretion curtain and consequently, different fractions of the X-ray emission from the magnetic poles pass through the accretion curtains and undergo absorption, producing the observed phase-dependent variation in the covering fraction and, hence, the spin modulation. Although a contribution from self-occultation cannot be excluded, the observed energy dependence of the spin modulation and associated HR variability suggest that it is not the dominant mechanism. A changing aspect of the blackbody-emitting region would primarily affect its projected area and hence its observed flux, but it would not be expected to produce HR variations unless significant temperature gradients were present across the small emitting spot, which are not detected.

\par In some IPs, asymmetric and extended accretion curtains can produce a slower reappearance compared to disappearance in the spin light curve (e.g. \citealt{vrielmann2005}).  To evaluate this possibility, we compared the phase intervals corresponding to the rise from pulse minimum to maximum and the subsequent decline back to minimum in the MOS and PN spin-folded light curves (see, Fig. \ref{fig:folded_tess_xmm}). The rise and decay intervals are comparable, differing by only $\sim$0.1 in phase, indicating that the X-ray spin-folded profiles of J0614 are approximately symmetric and do not exhibit the pronounced asymmetry expected from highly asymmetric accretion curtains. This suggests that the absorbing material is more uniformly distributed, rather than sharp transitions associated with highly asymmetric structures.

\subsection{Optical spin variability} \label{sec5.2}
We also detect modulation at the spin period in the optical bands, although with varying amplitude and more complex structure compared to the X-ray profiles. The optical spin modulation is likely produced by a combination of direct emission from the accretion curtains and reprocessed radiation from the inner accretion disc. As the WD rotates, the changing aspect of the accretion curtains and the varying illumination of the inner disc produce the observed optical pulsations at the spin period. The optical light curves exhibit broader and more structured profiles, with variations in pulse shape across different epochs. In particular, the profile in Sector 33 appears predominantly single-peaked, whereas those in Sectors 44 and 45 exhibit a prominent maximum with a double-humped structure, those in Sectors 43 and 71 show an asymmetric structure, and that in Sector 72 exhibits a broad, weakly structured maximum. The minima remain approximately aligned across all sectors, while the maxima vary significantly, indicating that the pulse-shape variability is primarily associated with changes in the emitting or reprocessing sites. The structured pulse profiles may also reflect harmonic contributions associated with reprocessing in different accretion regions. The double-humped appearance observed in Sectors 44 and 45, with the two peaks separated by a phase interval of about 0.4, could indicate a contribution from a larger accretion curtain or possibly from a secondary asymmetric pole. A much weaker indication of a secondary maximum is also present in the X-ray light curves, where a marginal excess appears roughly 0.5 in phase away from the primary maximum. During the \textit{XMM-Newton} epoch, the optical modulation observed with the OM is approximately in phase with the X-ray modulation, suggesting that both bands are dominated by emission associated with the same accreting magnetic pole. In contrast, the \textit{TESS} profiles appear approximately anti-phased with respect to the \textit{XMM-Newton} profiles, with the optical minima occurring close to the phase of the primary X-ray maximum, corresponding to an offset of roughly 0.5 spin cycle. The weak secondary X-ray maximum instead aligns more closely with the optical maxima observed in Sectors 33 and 72, while the remaining sectors exhibit broader and somewhat phase-shifted optical maxima. The observed phase offset may indicate that the relative contributions of the magnetic poles changed between the \textit{TESS} and \textit{XMM-Newton} epochs. In this scenario, the optical modulation detected by \textit{TESS} may be dominated by emission from a different magnetic pole, or by a different combination of pole emission and multiple reprocessing regions whose relative contributions and visibility vary with time, compared with the contemporaneous X-ray and optical modulation observed during the \textit{XMM-Newton} observation. Such changes in the accretion geometry could naturally account for the differences in pulse morphology and phase between the two epochs.

\subsection{Implications for accretion geometry} \label{sec5.3}
\begin{figure*}
\centering
\vspace{-0.2cm}
\includegraphics[width=16cm, height=4cm]{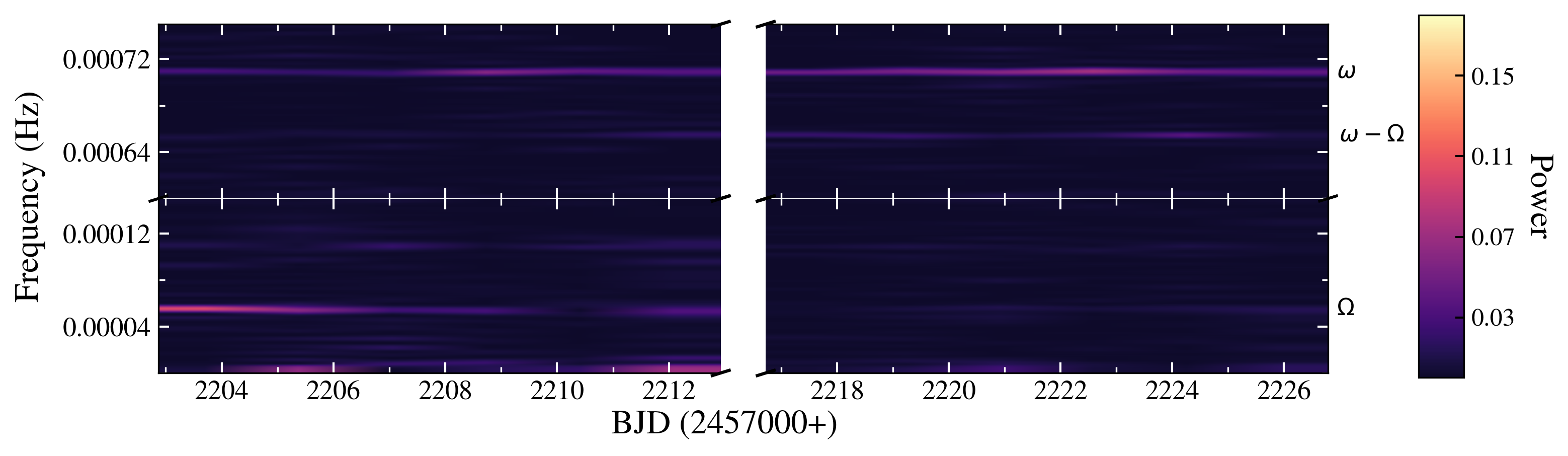}
\vspace{-0.2cm}
\includegraphics[width=16cm, height=4cm]{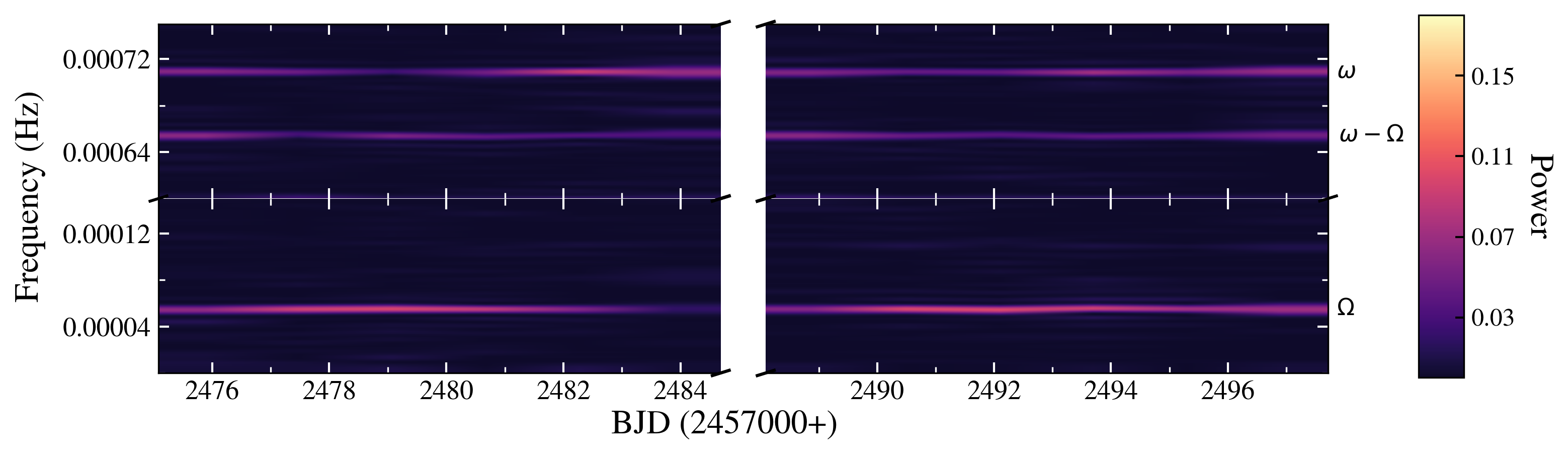}
\vspace{-0.2cm}
\includegraphics[width=16cm, height=4cm]{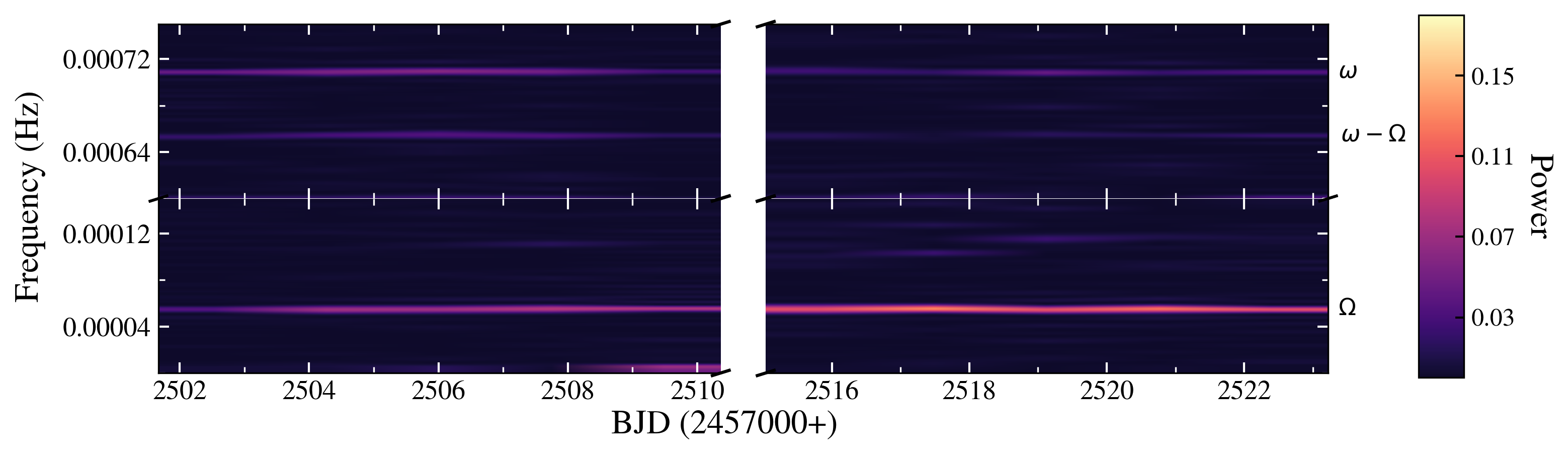}
\vspace{-0.2cm}
\includegraphics[width=16cm, height=4cm]{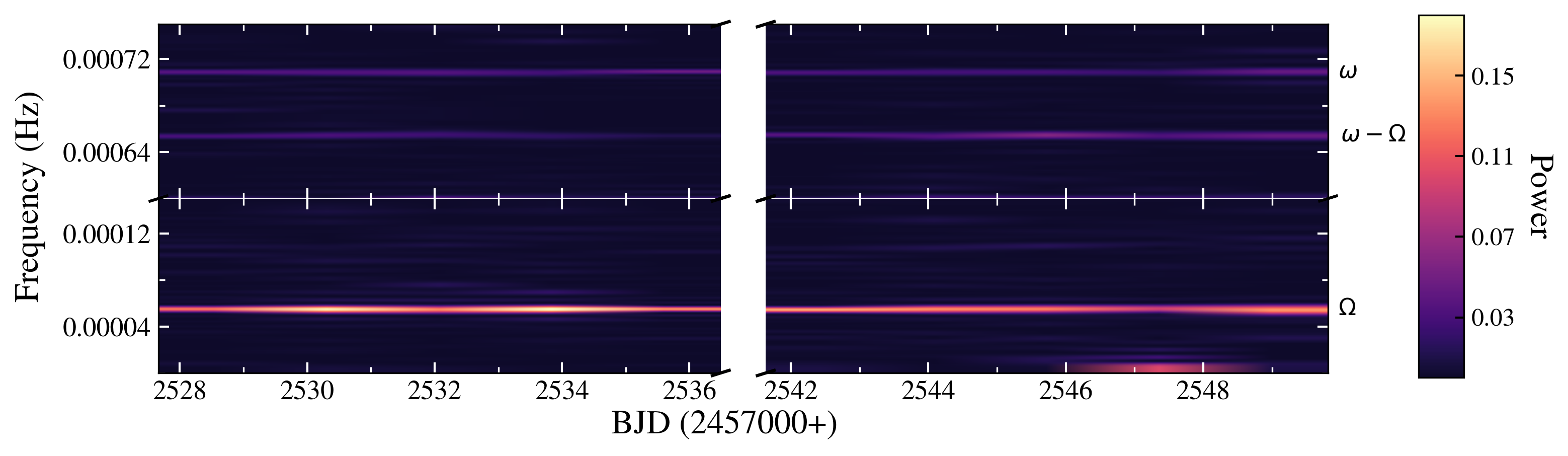}
\vspace{-0.2cm}
\includegraphics[width=16cm, height=4cm]{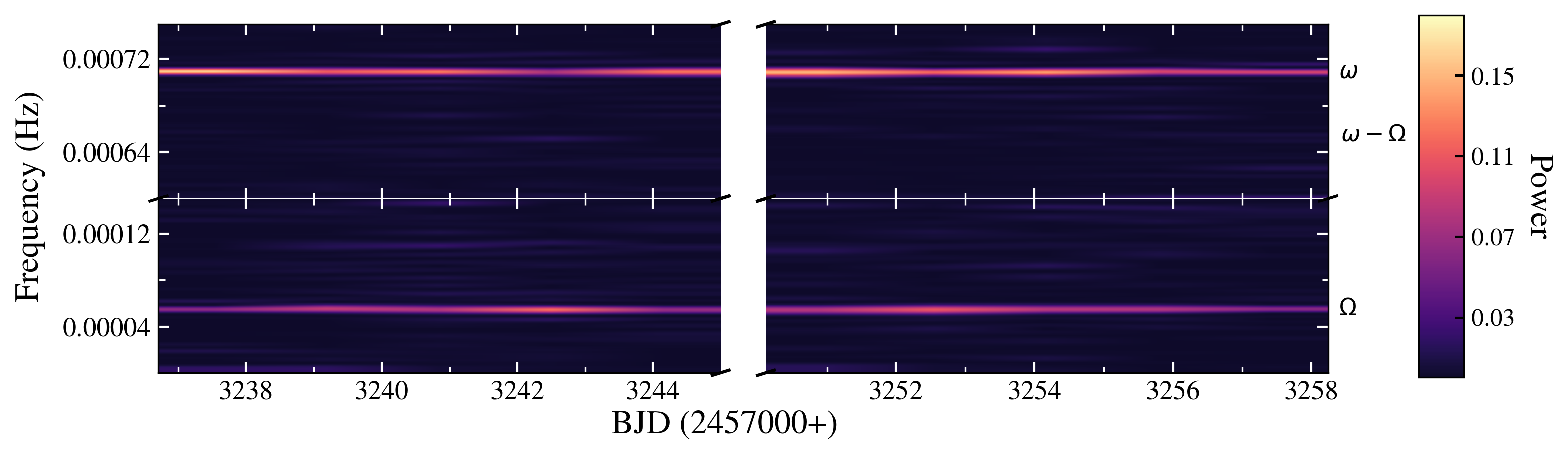}
\vspace{-0.2cm}
\includegraphics[width=16cm, height=4cm]{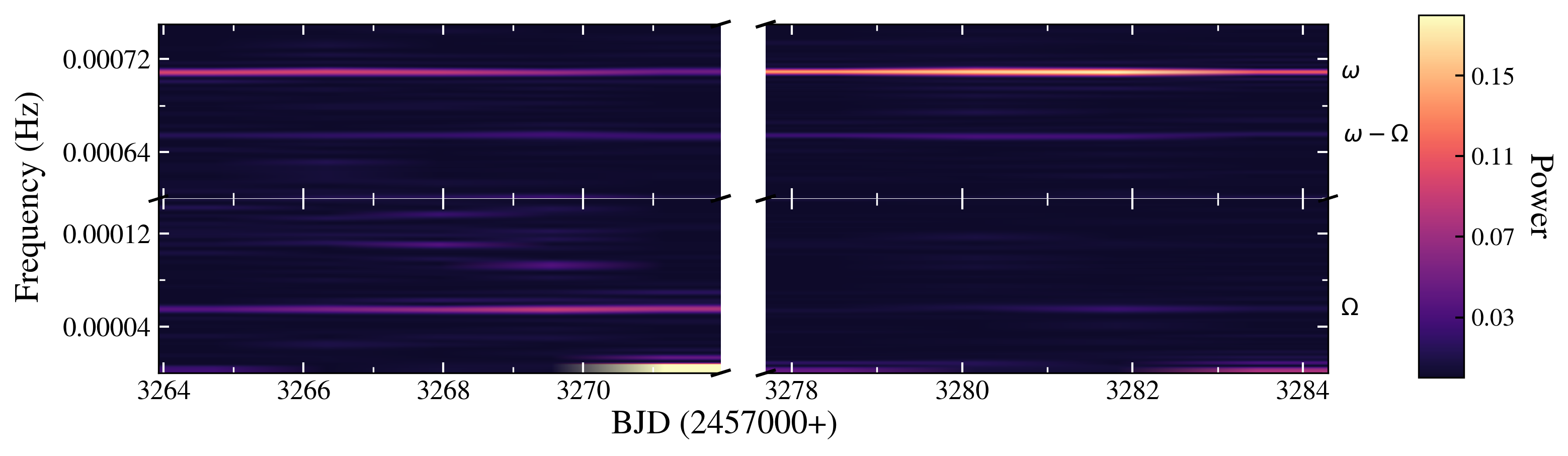}
\caption{Dynamical power spectra for \textit{TESS} Sectors 33, 43, 44, 45, 71, and 72. The panels are arranged from top to bottom in order of increasing sector number. In each panel, the upper half displays the frequency range centred on the spin and beat frequencies, while the lower half displays the frequency range centred on the orbital frequency. The colour bar indicates Lomb-Scargle power, and the same limits were adopted for all sectors. }
\label{fig:dps}
\end{figure*}

The power spectra behaviour reveals a clear distinction between the \textit{XMM-Newton} and \textit{TESS} observations. The simultaneous X-ray and OM data obtained during the \textit{XMM-Newton} epoch are dominated by the spin frequency, with no significant detection of the beat signal. In contrast, several \textit{TESS} sectors exhibit both the spin and beat frequencies, although the relative strength of both frequencies varies substantially between epochs. Sectors 33, 43, 44, and 72 show both the \s~and \be~frequencies, with the spin component clearly dominating the optical power spectra. Sector 45 appears to represent a transitional case in which the \s~and \be~frequencies display comparable power and amplitude, suggesting a temporarily enhanced contribution from the sideband-producing component. In contrast, Sector 71 shows no detectable beat frequency. In IPs, optical beat modulation is commonly attributed to the reprocessing of spin-modulated high-energy emission by structures fixed in the binary reference frame, such as the stream-disc impact region (the hotspot) or the secondary itself. The absence of the beat frequency in Sector 71 suggests that the corresponding reprocessing contribution was substantially weaker during this interval. We also detect the $\omega-2\Omega$ sideband in every sector except Sector 33. This frequency may originate either from modulation of the spin signal at the orbital harmonic (2$\Omega$) or from orbital modulation of the beat component. The dynamical power spectra further reveal that the beat signal exhibits substantial changes in strength between epochs, whereas the spin signal remains consistently detectable (see Fig. \ref{fig:dps}). Interestingly, during the second half of Sector 72, the spin signal becomes significantly stronger while the orbital signal weakens substantially compared to the first half of the sector. This behaviour suggests a significant change in the relative contributions of the spin- and orbital-modulated components on timescales of only a few days. The dominance of the \s~frequency in the X-ray data suggests that the system is predominantly disc-fed. The variable optical beat modulation observed in different \textit{TESS} epochs is therefore consistent with a changing contribution from the reprocessing regions responsible for producing the beat signal.

\par
\begin{figure}
\centering
\includegraphics[width=7cm, height=8cm]{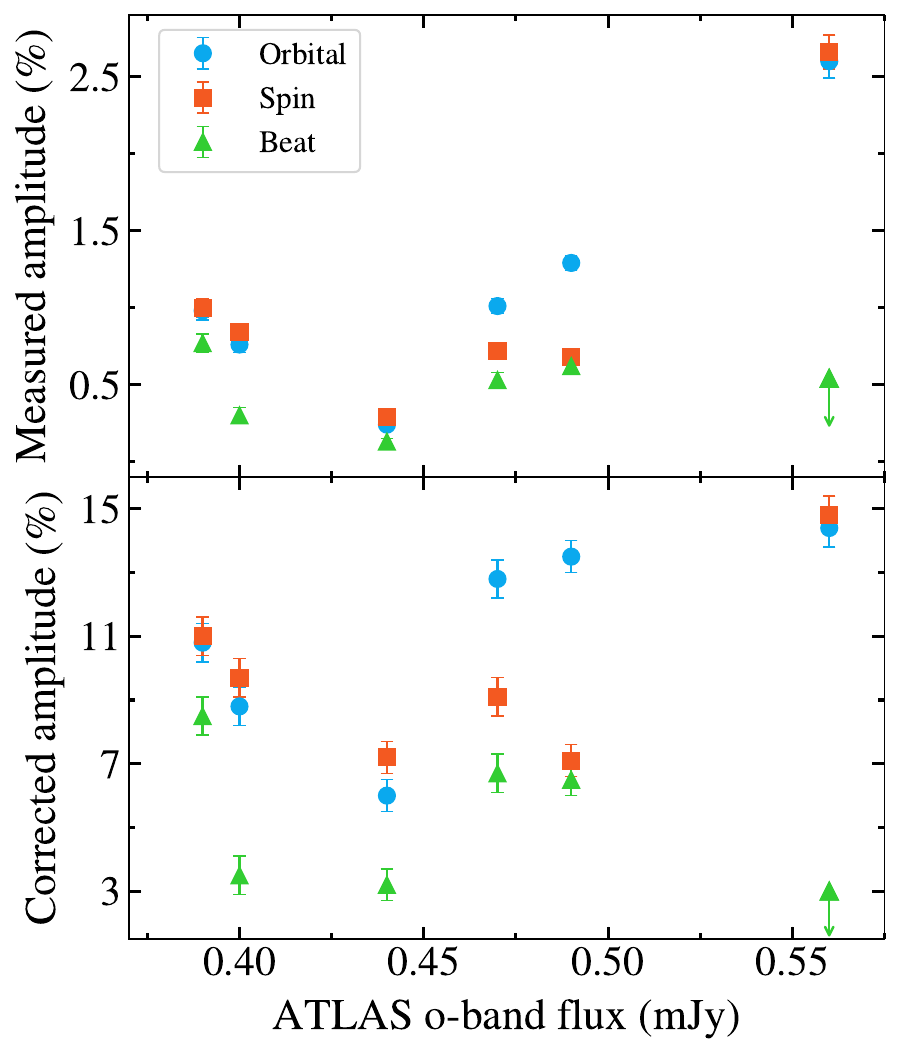} 
\caption{Measured and corrected amplitudes of the orbital, spin, and beat modulations as a function of ATLAS o-band flux. }
\label{fig:atlas_amplitude}
\end{figure}

 To further investigate whether these timing variations are related to changes in the brightness state of the system, we analysed quasi-simultaneous \textit{TESS} and ground-based ATLAS observations. Since fluxes from different \textit{TESS} sectors are not directly comparable, the ATLAS o-band photometry was used to trace long-term brightness variations in the system. The average magnitudes (fluxes in mJy) in Sectors 33, 43, and 72 lie in the range $\sim$17.31-17.45 (0.39-0.44), whereas Sectors 44 and 45 have average magnitudes of $\sim$17.19-17.25 (0.47-0.49), and Sector 71 reaches $\sim$17.05 (0.56), indicating that the system was in a comparatively brighter state during these epochs. Figure \ref{fig:atlas_amplitude} shows the relationships between the optical brightness of the system and both the measured and corrected modulation amplitudes. The brightest epoch (Sector 71) also exhibits the strongest orbital and spin modulations. Although no statistically significant correlations are found between the corrected modulation amplitudes and the optical brightness, the orbital modulation shows the strongest tendency to increase towards brighter optical states.  The beat amplitude does not show a clear correlation with system brightness, suggesting that the beat modulation amplitude is not solely affected by the global accretion state.  This lack of a clear correlation could reflect changes in the contribution of the reprocessing region(s), as also indicated by the changes in the phase of the beat modulation profiles observed across the different sectors (see Fig. \ref{fig:folded_beat}).

\par Because only one epoch combines relatively large spin and orbital amplitudes with a comparatively bright optical state, any physical interpretation remains tentative. Nevertheless, the observed behaviour is consistent with brighter states corresponding to enhanced mass transfer through the disc, producing stronger emission from the hotspot and therefore a stronger modulation at the orbital period. The spin modulation may likewise strengthen through enhanced reprocessing within the accretion flow, leading to a stronger optical spin signal during brighter states. Future observations spanning a wider range of brightness states will be essential to establish whether these apparent trends are robust.

\subsection{The post-shock region} \label{sec5.4}
The X-ray spectral properties of J0614 indicate a composite emission structure. The hard X-ray emission is consistent with optically thin plasma originating in the post-shock region, with a characteristic temperature of $11.8_{-1.5}^{+3.1}$ keV. In addition, a soft X-ray blackbody component at $62.8_{-6.3}^{+7.4}$ eV is required, which is more plausibly attributed to reprocessed emission from the heated WD surface. The absorption properties further indicate a complex local environment. In addition to a fully covering absorber with N$_{\rm H}$ of $\sim$1.0 $\times$ 10$^{22}$ cm$^{-2}$, consistent with the interstellar value along the line of sight ($6.2\times10^{21}$ cm$^{-2}$), the spectral fits require a partially covering absorber with N$_{\rm H}$ of $\sim$8.5 $\times$ 10$^{22}$ cm$^{-2}$, covering approximately 73\% of the X-ray source. The column density of the partial-covering absorber is consistent with the values found in the majority of IP systems \citep{2017PASP..129f2001M}. The observed 6.4 keV Fe K$\alpha$ emission is consistent with contributions of reflection from the WD surface and fluorescence in the pre-shock accretion column \citep{1999ApJS..120..277E}, but the present data do not allow these contributions to be disentangled.

\par The soft X-ray blackbody component in IPs spans a wide range of temperatures \citep[30–100 eV;][]{anzolin2008}. The most recent census indicates that `soft IPs' constitute approximately 30\% of the known population \citep[][and references therein]{2020AdSpR..66.1209D}. J0614 can therefore be classified as a member of this subgroup. Such hot blackbodies are thought to arise from small cores of the accretion spots at the WD poles. Using the normalisation of the blackbody component, we estimate an emitting radius of $154^{+231}_{-91}$ km. While the WD mass is poorly constrained, preventing a reliable estimate of the fractional emitting area, the inferred radius is much smaller than the expected WD radius, consistent with emission arising from a small, localised accretion footprint. For comparison, using the normalisation of the \texttt{apec} component, the Gaia distance, and assuming a post-shock electron density of  $\sim$10$^{16}$ cm$^{−3}$, we estimate an equivalent spherical radius of the post-shock plasma of $\sim$670 km. Although this estimate depends on the assumed electron density and simplified geometry, the inferred characteristic plasma size is of the same order of magnitude as the blackbody-emitting radius. Phase-resolved spectroscopy shows no statistically significant variation in the normalisation of the blackbody component. However, the relatively large uncertainties preclude a definitive statement about the amplitude of its spin modulation.

\subsection{Comparison with other IPs} \label{sec5.5}
The hard X-ray luminosities of most IPs are typically $\gtrsim$10$^{33}$ erg s$^{-1}$, while their bolometric X-ray luminosities are expected to be higher by a factor of a few. With an unabsorbed bolometric luminosity of $\approx$5.4$\times$10$^{34}$ erg s$^{-1}$, J0614 belongs to a small subset of IPs: only 7 out of the 67 confirmed systems reported in \citet{mukai2023} exhibit either hard or soft X-ray luminosities exceeding 10$^{34}$ erg s$^{-1}$. \citet{mukai2023} also showed that the X-ray luminosity and optical brightness, quantified through the disc brightness index (DBI), are strongly correlated, albeit with some outliers. Both quantities are therefore likely to be predominantly governed by a single parameter, most plausibly the mass accretion rate. We derived the DBI for J0614 following the standard method described in \citet{mukai2023}, obtaining a value of 0.63. With this DBI and an X-ray luminosity of 10$^{34}$ erg s$^{-1}$, J0614 is firmly located within the regime of optically defined high-luminosity IPs identified in that study.

\par Intermediate polars are observed to cluster around \ps/\po $\sim$0.1 in the orbital–spin period plane \citep{2004ApJ...614..349N}.  With \ps/\po $\sim$0.08, J0614 can be classified as an intermediate rotator. In this regime, accretion is generally expected to occur via a truncated accretion disc, although the exact accretion geometry also depends on the mass ratio of the system \citep{2004ApJ...614..349N}. J0614 exhibits strong X-ray spin pulsations, consistent with the behaviour of most IPs. However, several systems that are dominated by spin-modulated variability in X-rays also display beat-dominated modulations in the optical band (see e.g. \citealt{bernardini2012, 2017MNRAS.470.4815B}). In some cases, systems have also shown variability between spin- and beat-dominated signals in X-rays, reflecting changes in the dominant accretion mode. Such behaviour has been observed in a number of well-studied systems, including TX Col \citep{1997MNRAS.289..362N}, MU Cam \citep{staude2008}, FO Aqr \citep{evans2004, littlefield2020}, and V2400 Oph \citep{joshi2019}. Recent high-cadence observations from \textit{TESS} have further demonstrated that disc-overflow accretion can be variable in time. This behaviour has been identified in several systems, including TX Col \citep{2021ApJ...912...78R}, V902 Mon and SWIFT J0746.3–1608 \citep{2022MNRAS.512.6054R}, V709 Cas \citep{rao2026}, and Swift J2113.5+5422 \citep{sundriyal2026}. In the case of J0614, the absence of simultaneous spin and beat frequencies in X-rays argues against a disc-overflow accretion geometry during the observed epoch. However, the pronounced sector-to-sector variability in optical beat frequency suggests that the relative contribution of reprocessed emission varies on month-to-year timescales. Simultaneous high-cadence optical and X-ray observations will therefore be crucial for tracing the temporal evolution of the accretion flow in the system.

\section{Conclusions} \label{sec6}
Based on the pointed \textit{XMM-Newton} observation and complementary optical data from \textit{TESS}, we confirm J0614 as an IP and summarise the main results as follows:

\begin{itemize}
    \item We detect a coherent modulation at the WD spin period of $\sim$1411 s, consistent with previous optical determinations. The X-ray power spectrum is dominated by the spin frequency, supporting a predominantly disc-fed accretion geometry during the observed epoch.

    \item The spin modulation shows a strong energy dependence, and the hardness ratios vary in anti-phase with the intensity profiles. Phase-resolved spectroscopy shows significant changes in the covering fraction of the partial absorber across the spin cycle, indicating that the X-ray variability is primarily driven by absorption in the accretion curtains.

    \item The broadband X-ray spectrum is well described by a partially absorbed thermal plasma model with a plasma temperature of $\sim$12 keV, together with a soft blackbody component at $\sim$63 eV.
    
    \item The optical \textit{TESS} data reveal substantial sector-to-sector changes in the amplitudes of the orbital and spin modulations. The beat signal is absent in Sector 71, suggesting temporal variability in the contribution of reprocessed emission responsible for the beat modulation.

    \item The optical spin modulation exhibits an approximately 0.5-cycle phase offset between the \textit{TESS} and \textit{XMM-Newton} epochs, suggesting temporal changes in the dominant accreting magnetic pole and/or in the relative contributions of the associated reprocessing regions.

  \item The orbital and spin modulation amplitudes show a tendency to increase towards brighter optical states. Although no statistically significant correlation is found, this trend may be consistent with enhanced mass transfer.

    \item The long-term optical coverage reveals recurrent burst-like brightening events, adding J0614 to the growing group of IPs exhibiting short-duration optical bursts.
    
\end{itemize}
   
\begin{acknowledgements}
We thank the anonymous referee for the useful comments and suggestions, which have significantly improved the quality of the paper. This work is based on observations obtained with \textit{XMM-Newton}, an ESA science mission with instruments and contributions directly funded by ESA Member States and NASA. This paper includes data collected with the \textit{TESS} mission, obtained from the MAST data archive at the Space Telescope Science Institute (STScI). Funding for the \textit{TESS} mission is provided by the NASA Explorer Program. This work is based on the research supported in part by the National Research Foundation of South Africa. DdM and NM thankfully acknowledge financial support from INAF. ZAI gratefully acknowledges support from the UK Research and Innovation's Science and Technology Facilities Council (STFC) grant ST/X508767/1. During the final stages of preparation of this manuscript, \citet{Yan_2026} published an independent analysis of the XMM-Newton observation of J0614.   We gratefully acknowledge the contributions of the AAVSO observer community, whose photometric data and metadata resources were used in this study and made available through the AAVSO’s scientific archives. This work has made use of data from the Asteroid Terrestrial-impact Last Alert System (ATLAS) project. The Asteroid Terrestrial-impact Last Alert System (ATLAS) project is primarily funded to search for near earth asteroids through NASA grants NN12AR55G, 80NSSC18K0284, and 80NSSC18K1575; byproducts of the NEO search include images and catalogs from the survey area. This work was partially funded by Kepler/K2 grant J1944/80NSSC19K0112 and HST GO-15889, and STFC grants ST/T000198/1 and ST/S006109/1. The ATLAS science products have been made possible through the contributions of the University of Hawaii Institute for Astronomy, the Queen’s University Belfast, the Space Telescope Science Institute, the South African Astronomical Observatory, and The Millennium Institute of Astrophysics (MAS), Chile.  This work presents results from the European Space Agency (ESA) space mission Gaia. Gaia data are being processed by the Gaia Data Processing and Analysis Consortium (DPAC). Funding for the DPAC is provided by national institutions, in particular the institutions participating in the Gaia MultiLateral Agreement (MLA). The Gaia mission website is https://www.cosmos.esa.int/gaia. The Gaia archive website is https://archives.esac.esa.int/gaia.
\end{acknowledgements}

\bibliographystyle{aa}
\bibliography{ref.bib}

\clearpage
\onecolumn
\begin{appendix}
\section{Periodicities identified in \textit{TESS} data}
\begin{table}[ht!]
{\centering
\caption{Periods corresponding to the dominant peaks in the power spectra of J0614 of all six sectors.} \label{tab:periods_tess}
\renewcommand{\arraystretch}{1.4}
\begin{tabular} {lcccccc}
\hline
\multirow{2}{*}{Period} & \multicolumn{6}{c}{Sector}\\
\cline{2-7} & 33 & 43 & 44 & 45 & 71 & 72 \\
\hline
\po ~(h) & 5.09 $\pm$ 0.01 & 5.09 $\pm$ 0.01 & 5.09 $\pm$ 0.01 & 5.10 $\pm$ 0.01 & 5.10 $\pm$ 0.01 & 5.09 $\pm$ 0.01\\
\ptwoo ~(h) & 2.545 $\pm$ 0.004 & 2.546 $\pm$ 0.003 & --- & 2.548 $\pm$ 0.003 & 2.550 $\pm$ 0.007  & 2.551 $\pm$ 0.003 \\
\ps ~(s) & 1411.7 $\pm$ 0.3 & 1411.7 $\pm$ 0.3 & 1411.5 $\pm$ 0.3 & 1411.9 $\pm$ 0.3 & 1411.7 $\pm$ 0.3 & 1411.9 $\pm$ 0.3\\
\ptwos ~(s) & --- & 705.86 $\pm$ 0.07 & 705.89 $\pm$ 0.08 & 705.88 $\pm$ 0.07 & 705.87 $\pm$ 0.06 & 705.72 $\pm$ 0.08 \\
\pb ~(s)& 1529.2 $\pm$ 0.4 & 1529.5 $\pm$ 0.4 & 1529.6 $\pm$ 0.3  & 1529.4 $\pm$ 0.3 & --- & 1529.0 $\pm$ 0.4 \\
\ptwob ~(s) & --- & 764.70 $\pm$ 0.10 & --- & --- &   --- & ---  \\
\pbplus ~(s) & 1310.9 $\pm$ 0.2 & 1310.8 $\pm$ 0.2 & 1310.6 $\pm$ 0.2 & 1310.9 $\pm$ 0.3 & --- & ---\\
\pone ~(s) & --- & 1668.6 $\pm$ 0.4  & 1669.0 $\pm$ 0.4 & 1668.5 $\pm$ 0.4 & 1668.6 $\pm$ 0.4 & 1667.9 $\pm$ 0.3\\
\pthree ~(s) & --- & --- & 734.30 $\pm$ 0.09 & --- & --- & --- \\
\pfour ~(s) & --- & 679.71 $\pm$ 0.06 &  --- & --- & --- & --- \\
\hline          
\end{tabular}\par}
\vspace{3mm}
\hspace{3 mm}
\footnotesize
{\centering
\hspace{10 mm} \textbf{Notes.} Period uncertainties were derived from the Gaussian 1$\sigma$ widths of the fitted frequency peaks.}
\end{table}
\end{appendix}
\end{document}